\documentclass[aps,10pt,a4paper,amsmath,amssymb,nofootinbib,notitlepage,tightenlines,longbibliography,twocolumn,superscriptaddress]{revtex4-2}
\pdfoutput=1

\usepackage{natbib}
\usepackage{hyperref}
\usepackage{cleveref}

\usepackage{enumitem}

\usepackage{graphicx}
\usepackage[font={small,it}]{caption}
\usepackage{subcaption}
\newcommand{\caphead}[1]{{\bf #1}}
\usepackage{multirow}

\usepackage{bm}
\usepackage{bbm}

\usepackage{amsmath}
\usepackage{amsthm}
\usepackage{amssymb}

\newcounter{tlc}

\newtheorem{lemma}[tlc]{Lemma}
\crefname{tlc}{}{}

\newtheorem{definition}{Definition}
\newtheorem*{theorem*}{Theorem}
\newtheorem{theorem}{Theorem}

\newtheorem*{corollary*}{Corollary}

\usepackage{enumitem}
\allowdisplaybreaks[1] 
\usepackage[super]{nth}

\newcommand{\id}{\mathbbm{1}}

\makeatletter
\DeclareRobustCommand{\cev}[1]{%
  \mathpalette\do@cev{#1}%
}
\newcommand{\do@cev}[2]{%
  \fix@cev{#1}{+}%
  \reflectbox{$\m@th#1\vec{\reflectbox{$\fix@cev{#1}{-}\m@th#1#2\fix@cev{#1}{+}$}}$}%
  \fix@cev{#1}{-}%
}
\newcommand{\fix@cev}[2]{%
  \ifx#1\displaystyle
    \mkern#23mu
  \else
    \ifx#1\textstyle
      \mkern#23mu
    \else
      \ifx#1\scriptstyle
        \mkern#22mu
      \else
        \mkern#22mu
      \fi
    \fi
  \fi
}
\makeatother

\usepackage{stackengine}
\def\vecsign{\mathchar"017E}
\def\dvecsign{\smash{\stackon[-2.25pt]{\vecsign}{\rotatebox{180}{$\vecsign$}}}}
\def\dvec#1{\def\useanchorwidth{T}\stackon[-4.2pt]{#1}{\,\dvecsign}}
\stackMath

\newcommand{\past}[1]{\cev{#1}}
\newcommand{\future}[1]{\vec{#1}}
\newcommand{\pastfuture}[1]{\dvec{#1}}

\makeatletter
\newcommand*{\balancecolsandclearpage}{%
  \close@column@grid
  \clearpage
  \twocolumngrid
}
\makeatother

\newcommand{\kB}{k_\mathrm{B}}

\providecommand{\Pr}{}
\renewcommand{\Pr}[1]{\mathrm{P}\small(#1\small)}
\newcommand{\cPr}[2]{\mathrm{P}\small(#1 \,|\, #2\small)}
\newcommand{\Info}[2]{{I}\small(#1 \,; #2\small)}
\newcommand{\Ent}[1]{{H}\small(#1\small)}
\newcommand{\cEnt}[2]{{H}\small(#1 \,|\, #2\small)}
\newcommand{\cInfo}[3]{{I}\small(#1 \,; #2 \,|\, #3\small)}
\newcommand{\mvcInfo}[4]{{I}\small(#1 \,; #2 \,; #3 \,|\, #4\small)}
\newcommand{\insaneInfo}[5]{{I}\small(#1 \,; #2 \,; #3 \,; #4 \,|\, #5\small)}
\newcommand{\iInfo}[3]{{I}\small(#1 \,; #2 \,; #3\small)}

\newcommand{\ints}{\mathbb{Z}}

\newcommand{\NTU}{School of Physical and Mathematical Sciences, Nanyang Technological University,\\ 21 Nanyang Link, 637371, Singapore}
\newcommand{\IQOQI}{Institute for Quantum Optics and Quantum Information,\\ Austrian Academy of Sciences, Boltzmanngasse 3, A-1090 Vienna, Austria}

\begin{document} 

\title{The fundamental thermodynamic bounds on finite models}
\author{Andrew J.\ P.\ Garner}
\affiliation{\IQOQI}
\affiliation{\NTU}
\date{\today}
 
\begin{abstract}
The minimum heat cost of computation is subject to bounds arising from Landauer's principle.
Here, I derive bounds on finite modelling -- the production or anticipation of patterns (time-series data) -- by devices that model the pattern in a piecewise manner and are equipped with a finite amount of memory.
When producing a pattern, I show that the minimum dissipation is proportional to the information in the model's memory about the pattern's history that never manifests in the device's future behaviour and must be expunged from memory.
I provide a general construction of model that allow this dissipation to be reduced to zero.
By also considering devices that consume, or effect arbitrary changes on a pattern, I discuss how these finite models can form an information reservoir framework consistent with the second law of thermodynamics.
\end{abstract}

\maketitle

\section{Introduction}
\begin{figure}[t!]
\includegraphics[width=0.4\textwidth]{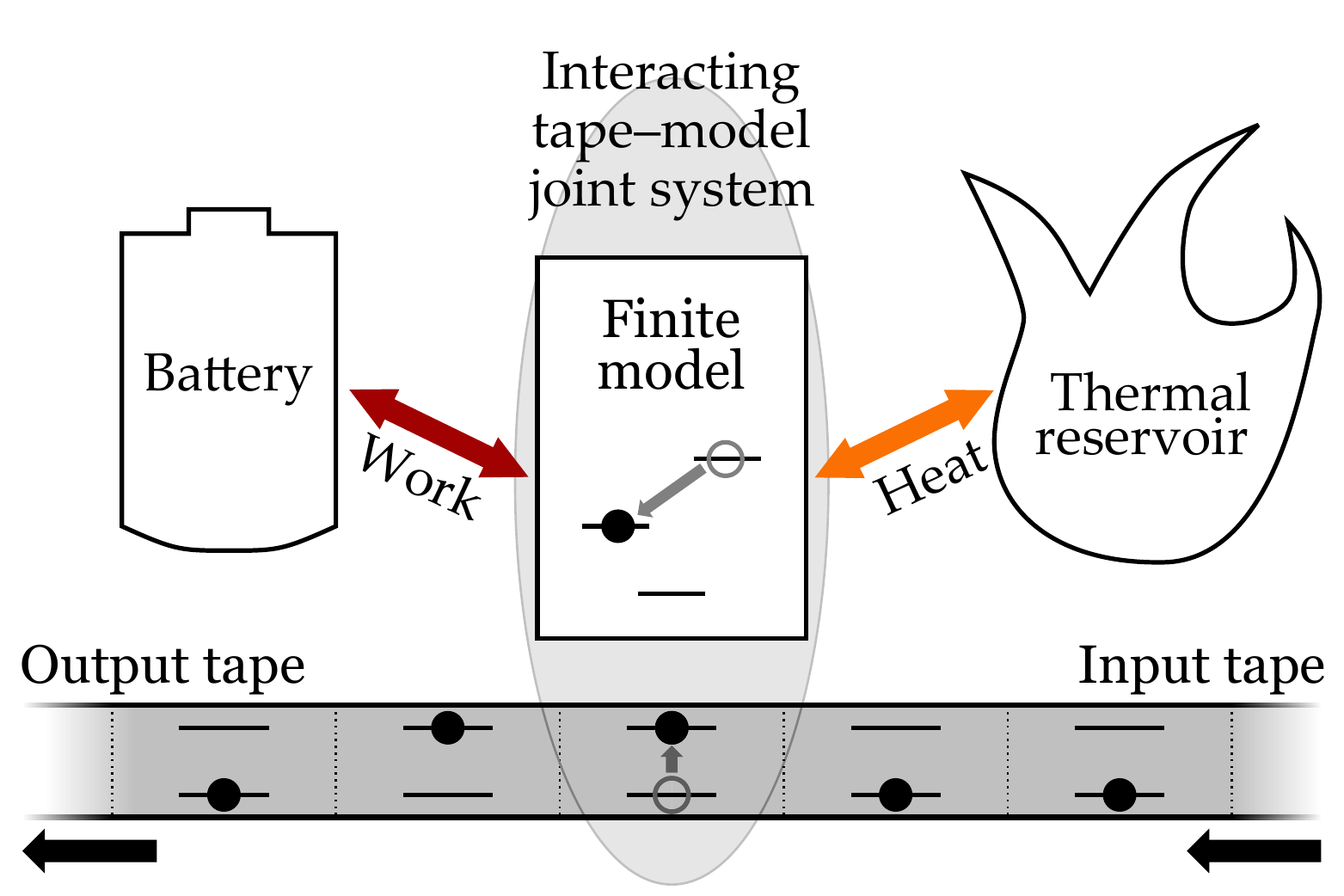}
\caption{
\label{fig:PatternManip}
\caphead{Thermodynamics of pattern manipulation.}
A series of configurable systems -- a tape -- passes through a {\em model} equipped with some internal memory.
At each time step, the model systematically interacts with the system on the tape, reconfiguring the tape and its internal memory.
To satisfy thermodynamic laws, the interaction may exchange work with a battery and heat with a thermal reservoir.
}
\vspace{-1em}
\end{figure}

Modern thermodynamics addresses the physical consequences of manipulating information~\cite{Landauer61,Bennett82}.
Before one reaches implementation--specific physical considerations (e.g.\ dissipation from internal resistance in transistors)
 there is a hierarchy of information-theoretical bounds.
These bounds arise from constraints, such as specifying the particular computational task performed, or limiting on the extent of information that can be accessed by the computer at any given time.
Here, I will consider specifically {\em finite models}: that is, the storage of information in a computer's memory about a pattern (i.e.\ discrete time-series data) that is used to anticipate or produce a pattern.
In this context, finite means that the task is performed in a piecewise manner (e.g.\ generating the sequence one step at a time), and the computation is done using only a finite amount of memory (see \cref{fig:PatternManip}).
Such finite models permeate the physical and quantitative sciences: from enzymes acting to copy DNA one a base--pair at a time, to meteorological supercomputers that forecast upcoming weather hour--by--hour.
Here, I will quantify the fundamental thermal limits on the tasks of pattern anticipation and pattern generation, 
 as given by the information-theoretical relationships between the model memory and the pattern.

There are two broad approaches to small-scale thermodynamics.
The first is from the ground up: explicitly construct a device and calculate its particular microscopic behaviour (e.g.\ heat exchanges in information ratchets~\cite{MandalJ12,BoydMC16,BoydMC17_b}).
This has the advantage of relating informational behaviour to other physical phenomena, and allows for intuitive physical modelling.
The second approach is top-down: one determines from general principles (such as adherence to the second law) universal bounds for {\em any} device that implements a particular {\em operational behaviour}, defined in terms of inputs and outputs~\cite{Landauer61,GarnerTVG17,BoydMC18}.
This has the advantage of making universal statements that hold true, even when subsequently applied to new physical mechanisms.
In this paper, I shall mainly adopt the second approach. 

The thermodynamics of patterns has recently been studied in the context of {\em information reservoirs}~
\cite{WiesnerGRV12,MandalJ12,StillSBC12,DeffnerJ13,Strasberg15,BoydMC16,GarnerTVG17,BoydMC17,BoydMC17_b,BoydMC18,LuJ19}.
Here, ordered {\em patterns} are treated a source of free energy -- 
	namely, a source of low-entropy states whose degradation allows the completion of tasks (such as resetting a random bit) that would otherwise require an explicit investment of work from a battery.
If an entire pattern could be acted on simultaneously,
 its thermodynamic treatment would be almost trivial:
 assuming degeneracy of the initial and final Hamiltonians,
 application of Landauer's principle~\cite{Landauer61,Bennett82} to the pattern 
 shows that the minimum average heat dissipation is proportional to the change in Shannon entropy between the input and output.
Taking in a disordered sequence and making it more ordered costs work; vice-versa releases it.
When only a limited portion of the pattern can be accessed at once (as required by finite models),
 the treatment becomes significantly more complicated.
To correctly function continually,
 a finite device must maintain a {\em model} of that pattern in its memory.
This model memory is also subject to thermodynamic laws~\cite{WiesnerGRV12,GarnerTVG17,BoydMC18}.

In this article, I probe the thermodynamics of three classes of finite model:
 those that generate a pattern, those that anticipate and consume one, and those that simply ``follow along''.
I begin with a brief review of what it means to be a finite model (\cref{sec:FM}),
 and describe a framework by which such models function as thermodynamic machines (\cref{sec:ThermalMachines}).
I show that the minimum cost of generating a chunk of the pattern is proportional to the amount of {\em discarded cryptic information}: stored knowledge about the pattern's that was never manifest the pattern's future and must be expunged from memory (\cref{sec:DisGen}).
I provide a construction and mechanism for a finite--model that avoids this cost (\cref{sec:OracularGen}). 
By considering the other devices that consume (\cref{sec:Consumer}), or arbitrarily change (\cref{sec:Transducer}) a pattern,
 I evaluate what limits the second law of thermodynamics places on the choice of memory,
 and establish how the picture of communicating finite models can form a thermodynamically consistent information reservoir framework.
I conclude with a discussion on the relation with these bounds and the specific ``prediction'' scenario in \citet{StillSBC12} (\cref{sec:Forecast}), supporting and generalizing their claim that dissipation results from ``useless \mbox{nostalgia}''.
This article thus formalizes a thermodynamic limit on allowed types of memory in physically-realizable models, and identifies the root cause of thermal dissipation during generation.

\section{Setting}
\label{sec:FM}
\subsection{Patterns and stochastic processes.}
Patterns can be quantified using the mathematical language of {\em stochastic processes}.
Let $X_t$ be a random variable, encapsulating some random choice from the alphabet $\mathcal{X}$.
A {\em pattern} is then the bi-infinite sequence $\pastfuture{X} := \cdots X_{t-1} X_t X_{t+1} \cdots$.
For classical information (i.e.\ without quantum correlations), the same $\pastfuture{X}$ can represent a spatial pattern or a temporal processes.
Consider an array of configurable systems (e.g.\ the tape in \cref{fig:PatternManip}) indexed by $t\in\ints$, where for each system, its configuration can be associated with some value in $\mathcal{X}$.
Then, for object $t$ the system's configuration is modelled by random variable $X_t$ and the entire tape realizes the pattern $\pastfuture{X}$.
Conversely, we could consider the state $X_t$ of just one system sampled at a series of discrete times, labelled by $t$.
The entire statistical history and future of this system's state is also represented by $\pastfuture{X}$.

One can convert between spatial and temporal pictures.
Imagine a tape travelling through a machine, where state $X_t$ is under the tape-head at time $t$.
The whole tape is the spatial realization of $\pastfuture{X}$, whereas describing the sequence of symbols found the tape-head at time $t$ is a temporal process, also expressed by $\pastfuture{X}$.
Switching between these two pictures is crucial for our thermodynamic understanding. 
In particular, to apply Landauer's principle~\cite{Landauer61,Bennett82} on all relevant random variables, the spatial picture is conceptually simpler (as per \cref{fig:PatternManip}).
Conversely, most literature on the relationship between memory and patterns (e.g.~\cite{CrutchfieldY89,ShaliziC01,CrutchfieldEM09,EllisonMJCR11,MahoneyEJC11,BarnettC15}) is framed in terms of stochastic processes, 
 but the insights are equally applicible to the spatial case~\cite{CrutchfieldF97,SuenTGVG17}.

In this article, we shall restrict our discussion to {\em stationary patterns}, where the statistics of $\pastfuture{X}$ have no explicit dependence on the index $t$ (though there can still be correlations between $X_t$ and $X_{t'}$ for two different values $t$ and $t'$).
Under this assumption, we take $t=0$ to be the ``current'' step of a pattern (e.g.\ the element under the tape head of \cref{fig:PatternManip}) without loss of generality.
A finite word formed by concatenating $k$ consecutive steps of the pattern from $t=1$ to $t=k$ inclusive is written as $X_{1:k} := X_1\ldots X_k$.
Expressions of the form $f(\past{X})$ are a shorthand for the limit $\lim_{L\to\infty} f( X_{-L:0})$, 
 and likewise $f(\future{X}) := \lim_{L\to\infty} f(X_{1:L})$.
The two implied infinite sequences $\past{X} := \cdots X_{-1} X_0$ and $\future{X} := X_{1} X_{2} \ldots$ are known as the {\em past} and {\em future} of the pattern respectively.

In the context of pattern thermodynamics, the (unconditioned) entropy per symbol $\Ent{X_t}$ is less important than the pattern's {\em entropy rate}~\cite{Lindgren88} $h_{X} := \lim_{L\to\infty} \frac{1}{L} \Ent{X_{0:L-1}} \to \cEnt{X_0}{\past{X}}$.
This quantity represents the effective amount of new entropy per step, as viewed (e.g.)\ by an agent with access to the entire history of the pattern.
For independent and identically distributed (i.i.d.)\ patterns (i.e.\ without correlations between successive steps), then $h_X = \cEnt{X_0}{\past{X}}=\Ent{X_0}$.

\subsection{Finite models.}
The manipulation of information inevitably results in a reconfiguration of the physical system on which the physical information was encoded~\cite{Landauer91}.
As such, the change of one pattern $\pastfuture{X}$ into another $\pastfuture{Y}$,
 should be evaluated as a physical process.
This means there may be some physical limitations on the manner by which such a transformation can be performed.
Here, we will consider specifically {\em finite models}:
\begin{definition}
\label{def:FM}
A {\bf finite model} is a machine that manipulates a pattern such that:
\begin{enumerate}
\item It reads/writes a finite amount of the pattern at any time step (e.g.\ only has access to the part under the tape head in \cref{fig:PatternManip}). 
\item It has a finite amount of internal memory (and so does not become a version of Maxwell's demon by dumping old data into an arbitrarily large database~\cite{Landauer61,Bennett82}).
\item It can be repeatedly used to effect an arbitrarily large part of the transformation. 
\item It acts on the pattern, visiting each step once, in a pre-determined order.
\end{enumerate}
\end{definition}

Such a definition is conceptually close to the {\em transducers} discussed in~\citet{BarnettC15}, 
 but there is a particular distinction in emphasis:
Here, we will derive bounds for any {\em particular} information--theoretical choice of internal memory,
 whereas \citet{BarnettC15} seek to provide a systematic optimal description of input--output processes as a state machine, with this optimisation over all such choices of memory (and thus fixing one particular choice of memory).
To avoid confusion between these definitions\footnote{I also wish to avoid invoking the regularly imagined two-tape picture of  ``finite state transducers'' from automata theory -- since the thermodynamics are calculated in a ``one tape'' framework, like \cref{fig:PatternManip}.},
 I thus use the term ``finite model'' with the mere promise that it is some device satisfying the desiderata above.

However, we make no assumptions on the specific physical mechanism by which the finite model is implemented,
 instead deriving universal ``top--down'' information theoretic bounds.
For formalisms that realize \cref{def:FM} in a constructive (i.e.\ ``bottom-up'') manner, one could consult (e.g.)\ the trajectory formalism~\cite{Alicki79,AlickiHHH04,Aberg13} implementation in \citet{GarnerTVG17},
 the isothermal Markov channels in \citet{BoydMC18},
 or (for certain transductions) the fluctuation--theorem--inspired~\cite{Jarzynski97,Crooks99} approach of  information ratchets~\cite{MandalJ12,BoydMC16,BoydMC17_b}.
 
Here, we will focus on three sub-classes of finite model, classified by their operational behaviour. 
The first two I define here -- the third (a {\bf forecaster}) will be discussed in \cref{sec:Forecast}:
\begin{definition}
A {\bf generator} of $\pastfuture{Y}$ is a finite model that takes an i.i.d.\ sequence $\pastfuture{X}_{\rm dflt} := \ldots X_{\rm dflt} X_{\rm dflt} X_{\rm dflt}\ldots$, and configures it into the pattern $\pastfuture{Y}$.
\end{definition}
\begin{definition}
A {\bf consumer} of $\pastfuture{X}'$ is a finite model that takes a pattern $\pastfuture{X}'$, and resets it into the i.i.d. sequence $Y_{\rm dflt}' := \ldots Y'_{\rm dflt} Y'_{\rm dflt} Y'_{\rm dflt} \ldots$.
\end{definition}

\subsection{Model memory.}
\label{sec:mem}
Key to the calculation of thermodynamic bounds is
 the relationship between a finite model's internal memory (denoted $R$) and the involved patterns (denoted by $\future{Z}$ as a stand-in for $\pastfuture{Y}$ in the generator or $\pastfuture{X}$ in the consumer).

For generators, without loss of generality, we can consider memory that leverages {\em all} information available from the history of the pattern pertinent to its future statistics.
This is because our explicitly finite generators must produce a future $\future{Z}$ that has correct correlations with past $\past{Z}$, but with only $R$ as a proxy for $\past{Z}$,
 implying a Markov chain (see, e.g.,\ \citet{CoverT91}) $\past{Z} \to R \to \future{Z}$.
Consequently, the mutual information $\Info{\future{Z}}{\past{Z}{R}} = \Info{\future{Z}}{R}$,  implying a set of data-processing inequalities (\cref{app:DPI}),
 such that the memory $R$ acts as a ``causal shield'' between the pattern's past and future.
This requirement is in contrast to an information--bottleneck~\cite{TishbyPB99,Still14} approach, where the capacity of the model to perfectly produce the pattern can be limited.
Conversely, there is no such data--processing--motivated reason why a consumer should keep knowledge about its prior inputs -- but it has been shown that failing to do so incurs a thermodynamic penalty in its operation~\cite{BoydMC18}.

\begin{figure}[hbt]
\centering
\includegraphics[width=0.3\textwidth]{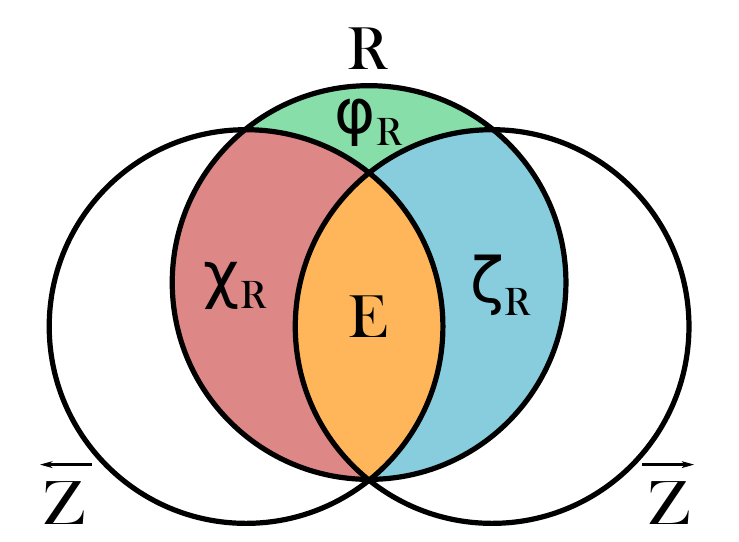}
\caption{
\label{fig:VennSimple}
\caphead{The information--theoretic relationships between a generator's memory $R$ and a pattern $\protect\pastfuture{Z}$.}
See \cref{app:InfoAnatomyWhole} for details.
Of particular thermodynamic interest in this article will be the cryptic information~\cite{MahoneyEJC11} $\chi_R$ and the oracular information~\cite{RuebeckJMC18} $\zeta_R$.
There is no region intersecting $\past{Z}$ and $\future{Z}$ but excluding $R$, since $\Info{\future{Z}}{\past{Z}{R}} = \Info{\future{Z}}{R}$.
}
\end{figure}

Computational mechanics provides us with the tools for classifying the information in such memory in terms of its relationship with a pattern~\cite{CrutchfieldEJM10,EllisonMJCR11} (see \cref{app:Anatomy}).
In particular, we can subdivide $\Info{\pastfuture{Z}}{R}$ (see \cref{fig:VennSimple}) into parts relating to the future of the pattern, the past of the pattern, or both.

\section{Thermodynamic bounds}
\subsection{Generators and consumers as thermal machines.}
\label{sec:ThermalMachines}

\begin{figure}[hbt]
\vspace{-2em}
\includegraphics[width=0.33\textwidth]{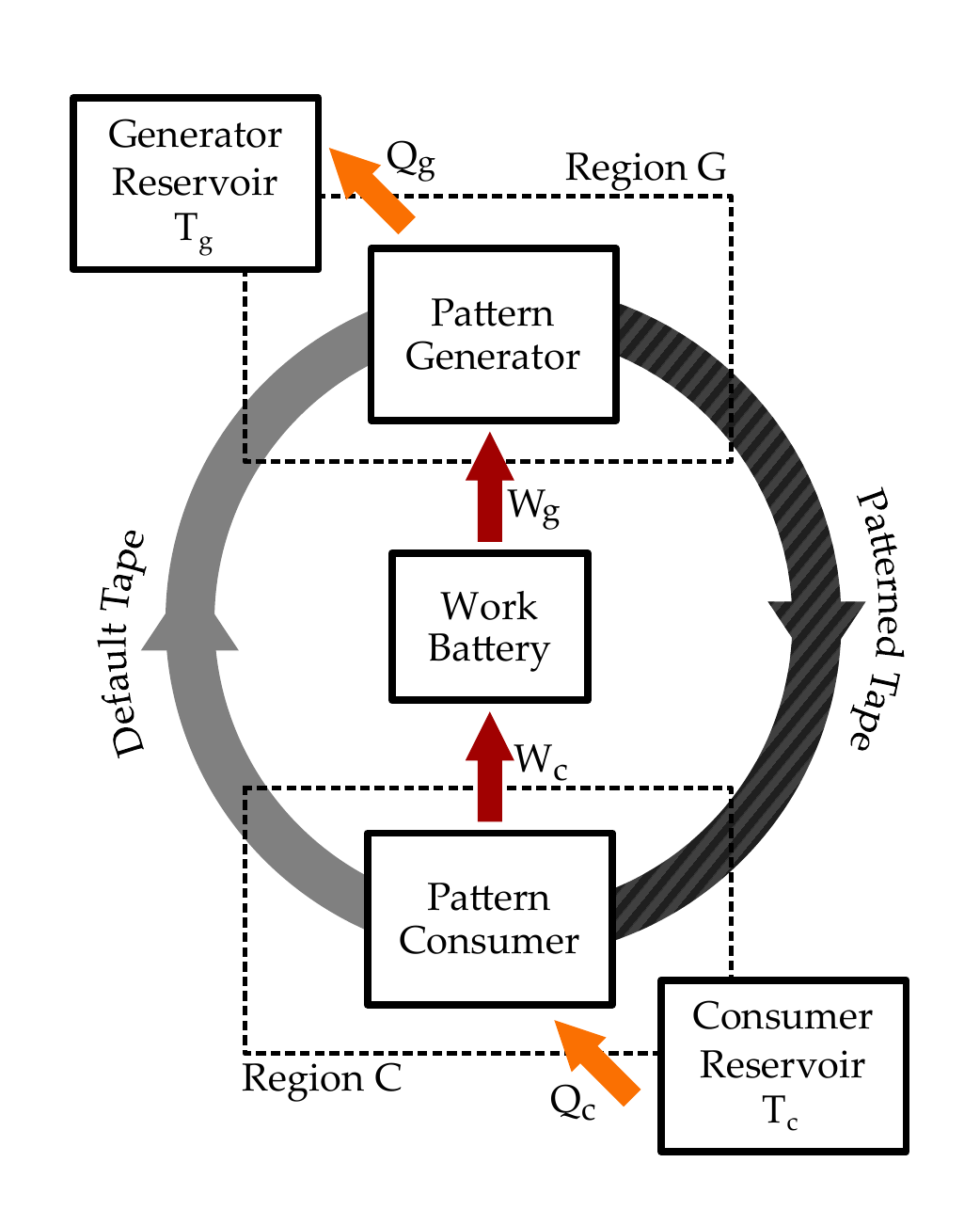}
\caption{
\label{fig:Carnot}
\caphead{Closed cycle of generation and consumption.}
A loop of tape circles through two machines.
The {\em generator}, configures the tape according to some pattern, perhaps requiring some input of work.
The {\em consumer}, anticipates the incoming pattern on the tape, and resets the tape back to its default state, perhaps extracting work in the process.
Each region in a dashed box corresponds to a setting like \cref{fig:PatternManip}.
}
\end{figure}

Let us examine how finite models can be employed in a thermal setting.
We shall consider cyclic behaviour (as in \cref{fig:Carnot}),
 where the output tape of the generator is then fed into the consumer, and vice versa.
In this configuration the generator produces exactly the pattern that the consumer is configured to consume,
 the i.i.d.\ ``default tape'' is likewise identical between the two,
 and the generator and consumer produce (resp.\ consume) the same number of steps $k$ of the pattern.
Due to the stationarity of the pattern and the manipulators (as per \cref{def:FM}), 
 the net macroscopic effect of such a cycle is encapsulated entirely by the exchanges between the work battery and heat reservoirs.
As such, by considering the entire system of the generator, consumer and the loop of connecting tape as a composite ``working medium'',
 then the second law upper bounds the efficiency with which the battery can be charged by the Carnot limit (i.e.,\ if temperatures $T_c\geq T_g$, the maximum effiency when operating as a heat--engine is $\eta = \frac{W_c-W_g}{Q_c} = 1 - \frac{T_g}{T_c}$).

To obtain tighter bounds (or to show that there is no information--theoretic reason to forbid reaching Carnot effiency), 
 we consider the system with more nuance,  
 adopting the information reservoir framework~\cite{WiesnerGRV12,MandalJ12,StillSBC12,DeffnerJ13,Strasberg15,BoydMC16,GarnerTVG17,BoydMC17,BoydMC18,LuJ19} to probe each of  \cref{fig:Carnot}'s dashed regions (i.e.\ treating them as instances of \cref{fig:PatternManip}).
Here, the alterations to the input and output tape can be treated akin to charging another type of battery (as we substantiate in the following).

\subsection{The work cost of pattern generators.}
\label{sec:DisGen}
Let us evaluate the bounds for a generator of pattern $\pastfuture{Y}$ (dashed region G of \cref{fig:Carnot}).
This device acts on words of length $k$ to transform them from the i.i.d.\ state ${X_{\rm dflt}}^{\otimes k}$ to the patterned word $Y_{1:k}$, and updates its internal memory from $R_0$ to $R_{k}$.
The total change in entropy of the length-$k$ tape section and memory is:
\begin{align}
\label{eq:EntGen}
\Delta H = \Ent{R_{k} Y_{1:k}} - \Ent{R_0 {X_{\rm dflt}}^{\otimes k}}.
\end{align}

Let us assume that that every microstate of the pattern and memory is equally energetically favourable (i.e.\ setting the initial and final Hamiltonian to zero).
Then, since the finite model only has access to this length-$k$ region of the tape,
 Landauer's principle~\cite{Landauer61} gives us the minimum work cost $W_g = -\kB T_g \Delta H$.
Rearranging \cref{eq:EntGen} (proven below): 
\begin{align}
\beta W_g & = k\left[\Ent{X_{\rm dflt}} - h_Y \right] \nonumber \\
& + \cEnt{R_0}{Y_{1:k} R_k} - \cEnt{R_{k}}{Y_{1:k} R_0} + \zeta_R\!\left(k\right).
\label{eq:OutputCost}
\end{align}
where $\beta:=1/\kB T$,  $h_Y := \cEnt{Y_1}{\past{Y}}$ is the entropy rate of the pattern $\pastfuture{Y}$,
 and 
\begin{align}
\zeta_R\!\left(k\right) := \cInfo{Y_{1:k}}{R_0}{\past{Y}}
\end{align}
 is the {\em oracular information}~\cite{RuebeckJMC18} that $R_0$ contains about the next word of length $k$\footnote{
 The {\em conditional mutual information} $\cInfo{A}{B}{C} := \Info{A}{B,C} - \Info{A}{C}$ encompasses the correlations between $A$ and $B$ that are not explained by $C$~\cite{CoverT91}.}.
In particular, the term $\zeta_R$ describes the additional information that the memory has about the output pattern that could not be inferred from the history of outputs thus far.

\begin{lemma}
\label{lem:Gen1}
Proof of above.
\begin{proof}
Expand $\Ent{R_0 Y_{1:k} R_{k}}$ in two different orders,
\begin{align}
\Ent{R_0 Y_{1:k} R_{k}}& = \Ent{R_0} + \cEnt{Y_{1:k}}{R_0} + \cEnt{R_{k}}{Y_{1:k} R_0} \nonumber\\
 & = \Ent{Y_{1:k}R_{k}} + \cEnt{R_0}{Y_{1:k} R_{k}} 
\end{align}
and thus re-express the first term of \cref{eq:EntGen} as
\begin{align}
\label{eq:firstTermEntGen}
\Ent{R_{k} Y_{1:k}} & = \Ent{R_0} + \cEnt{Y_{1:k}}{R_0} \nonumber \\
& \quad + \cEnt{R_{k}}{Y_{1:k} R_0} - \cEnt{R_0}{Y_{1:k} R_{k}}.
\end{align}
Now consider two expansions of  $\cEnt{Y_{1:k}}{\past{Y}}$:
\begin{align}
\label{eq:AppAExp}
\cEnt{Y_{1:k}}{\past{Y}}  & = \cEnt{Y_k}{\past{Y}Y_{1:k\!-\!1}} + \ldots + \cEnt{Y_1}{\past{Y}} \nonumber \\ 
&= k h_{Y}, \\
& = \cInfo{Y_{1:k}}{R_0}{\past{Y}} + \cEnt{Y_{1:k}}{\past{Y}R_0} \nonumber\\
&= \zeta_R\!\left(k\right) + \cEnt{Y_{1:k}}{R_0}.
\end{align}
The first expansion uses stationarity, and the definition of $h_{Y}$.
The second expansion uses the definitions of conditional mutual information and of $\zeta_R\!\left(k\right)$,
 then applies $\cEnt{Y_{1:k}}{\past{Y}R_0}=\cEnt{Y_{1:k}}{R_0}$.
This final step is possible because as a finite model generator, $R_0$ contains all the information required to generate $Y_{1:k}$  (see also \cref{app:DPI}).
Hence, we conclude: 
\begin{align}
\label{eq:OracularCorrection}
\cEnt{Y_{1:k}}{R_0} & = kh_Y - \zeta_R\!\left(k\right)
\end{align}

Since since all systems in the second term of \cref{eq:EntGen} are independent,
 it trivially expands as $\Ent{R_k {X_{\rm dflt}}^{\otimes k}} = \Ent{R_k} + k \Ent{X_{\rm dflt}} = \Ent{R_0} + k \Ent{X_{\rm dflt}}$ (where the final step uses stationarity).
Substituting this, and \cref{eq:OracularCorrection} (via \cref{eq:firstTermEntGen}) into \cref{eq:EntGen}, gives the claim.
\end{proof}
\end{lemma}

The first term of \cref{eq:OutputCost} is entirely independent of the particular choice of generator memory, and directly corresponds to the change in the tape's entropy rate.
As such, we can define a per-symbol tape work cost $W_{tape}$ that is a function only of the particular choice of pattern:
\begin{align}
\label{eq:tapecost}
\beta W_{tape} & := \left[\Ent{X_{\rm dflt}} - h_Y \right],
\end{align}
and a per-generation memory update cost $W_{\rm mem}^k$ that is a function of the pattern, the number of steps generated, and the choice of memory:
\begin{align}
\label{eq:GenDiss}
\beta W_{\rm mem}^k & := \cEnt{R_0}{Y_{1:k} R_{k}} - \cEnt{R_{k}}{Y_{1:k} R_0} + \zeta_R\!\left(k\right),
\end{align}
Thus, \cref{eq:OutputCost} can be alternatively expressed as $W_g = k W_{tape} + W_{\rm mem}^k$.

The expression $W_{\rm mem}^k$ has a similar form to Eq.~1 in \citet{GarnerTVG17}, but contains the extra term $\zeta_R$, 
 resulting from its derivation for a much more general class of model memory.
However, although $\zeta_R\!\left(k\right)\geq0$, the admission of oracular information allows the difference between the two other terms of \cref{eq:GenDiss} to be negative (which would otherwise not be possible~\cite{GarnerTVG17}).
$W_{\rm mem}^k$  can be further re-arranged to the main result of this article:
\begin{theorem}
\label{thm:CrypticDissipation}
For a finite model with memory $R$ that generates $k$ steps of a pattern $\pastfuture{Y}$ at a time,
 the extra minimum work cost due to this choice of memory is given 
\begin{align}
\label{eq:CrypticDissipation}
\beta W_{\rm mem}^k =  \cInfo{\past{Y}}{R_0}{\future{Y}R_{k}}.
\end{align}
\begin{proof}
The proof follows from Lemma~\ref{lem:Gen1}.
Rearranging \cref{eq:GenDiss} to the form of \cref{eq:CrypticDissipation} requires extensive technical framework,
 incorporating the following two features that I outline only in brief here (with full details in \cref{app:InfoAnatomyWhole,app:DPIZeros,app:ToG}):
{\bf (1)} $\past{Y} \to R_0 \to Y_{1:k} R_k$ must be a Markov chain
 (i.e.\ $\cPr{Y_{1:k} R_k}{R_0} = \cPr{Y_{1:k} R_k}{\past{Y} R_0}$), 
 so the data processing inequality sets many of the generator's possible information quantities to zero -- see \cref{app:DPIZeros}.
{\bf (2)} In stationary and continuous operation, the various classes of information (see \cref{fig:VennSimple}) that constitute a generator's knowledge about a pattern are independently conserved as the generator updates -- see \cref{app:ToG}.

From these features (with the visual aid of an information diagram -- see \cref{app:InfoAnatomyWhole}), ultimately one can equate the RHS of \cref{eq:GenDiss} with the RHS of \cref{eq:CrypticDissipation}.
\end{proof}
\end{theorem}

An immediate corollary is that $W_{\rm mem}^k\geq 0$, since bipartite conditional mutual informations are non-negative.
Thus, from information theory alone, we have the arithmetic bound ``$W_g \geq k W_{tape}$'' on the work cost over any choice of generator memory 
 -- though it remains to show (we do so in the next section) that this can be saturated by a particular memory choice.
 
The quantity on the RHS of \cref{eq:CrypticDissipation} it is the information stored in the memory at time $0$ about the history of the pattern that has nothing to do with the future of the pattern, and was subsequently ejected from the memory by time $k$.
In computational--mechanical language, this is the {\em discarded cryptic information}~\cite{MahoneyEJC11,MahoneyEJC11} (see \cref{app:Anatomy}) -- 
 and as we shall discuss later, is conceptually similar to the ``useless nostalgia'' of \citet{StillSBC12} (albeit in a very different setting). 
In this sense, a generator is inefficient when it the needs to clean up a useless record of the past.
 
Finally, we remark that $W_g= kW_{\rm tape} + W_{\rm mem}^k$ is an information--theoretic bound on the work cost derived tightly from Landauer's principle as applied to a particular choice of model memory.
There may well be additional details of implementation (such as energy--level manipulating protocols required to complete in a short amount of time) that introduce excess work dissipation above this limit.
However, there are many ``bottom-up'' frameworks~(e.g.\ \cite{AlickiHHH04}) that can derive the work exchange required to configure a system from dynamical origins (without direct recourse to Landauer's principle),
  and yet still saturate the bounds set by Landauer's limit.
For the remainder of this article we will thus make the assumption that for every choice of model memory, 
  there always exists an implementation that operates exactly as to saturate Landauer's bound (e.g.,\ the isothermal Markov channels described in \citet{BoydMC18}).

\subsection{Minimizing the work cost of generators.}
\label{sec:OracularGen}
With free choice of memory $R$, is there a systematic choice such that \cref{eq:CrypticDissipation} is minimized?
In \citet{GarnerTVG17} this minimum over the subset of models with no oracular information (i.e.\ $\zeta_R=0$) 
 was was found as the generator whose memory is in one-to-one correspondence with the {\em causal states}~\cite{CrutchfieldY89,ShaliziC01} of the generated pattern.
This amounts to storing the minimum statistically--relevant synopsis of the pattern's history,
 by recording the equivalence class of the relation~$\sim_\varepsilon$ partitioning the histories:
\begin{equation}
\past{x} \sim_\varepsilon \past{x}' \quad \mathrm{iff} \quad \cPr{\future{X}\!=\!\future{x}}{\past{X}\!=\!\past{x}} =  \cPr{\future{X}\!=\!\future{x}}{\past{X}\!=\!\past{x}'}  \; \forall \future{x}.
\end{equation}
A model whose memory exactly corresponds to the causal states (the above equivalence classes) is known as an {\em $\varepsilon$--machine}.
However, for general processes these  $\varepsilon$--machine generators have $W^{k}_{\rm mem} > 0$, since they typically contain cryptic information, 
 and hence are not thermodynamically optimal.

\begin{figure*}[tbh]
\begin{centering}
\begin{tabular}{ccc}
\begin{subfigure}[t]{0.3\textwidth}
\includegraphics[width=\textwidth]{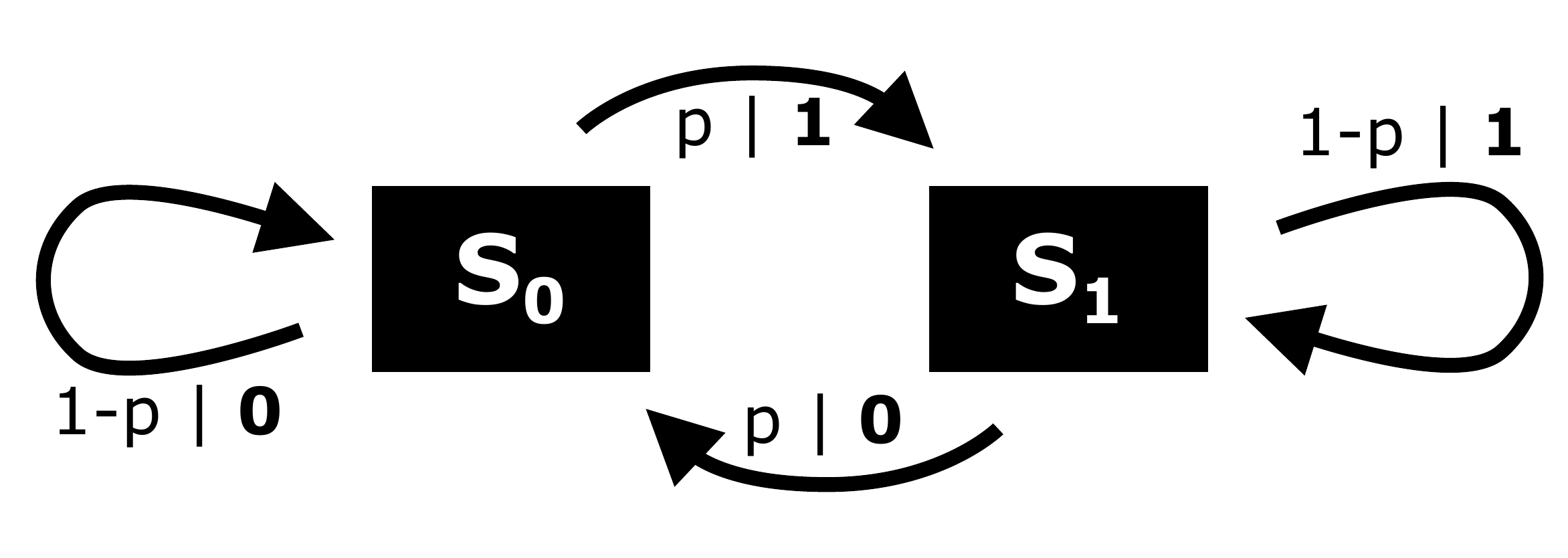}
\subcaption{\label{fig:PCeps} {\em $\varepsilon$--machine.}
The two causal states $s_0$ and $s_1$ correspond to whether the last output was respectively a $0$ or a $1$.
}
\end{subfigure}
&\hspace{1.5em}
&
\multirow{2}{*}[4.85em]{
\setcounter{subfigure}{2}
\begin{subfigure}[t]{0.3\textwidth}
\includegraphics[width=\textwidth]{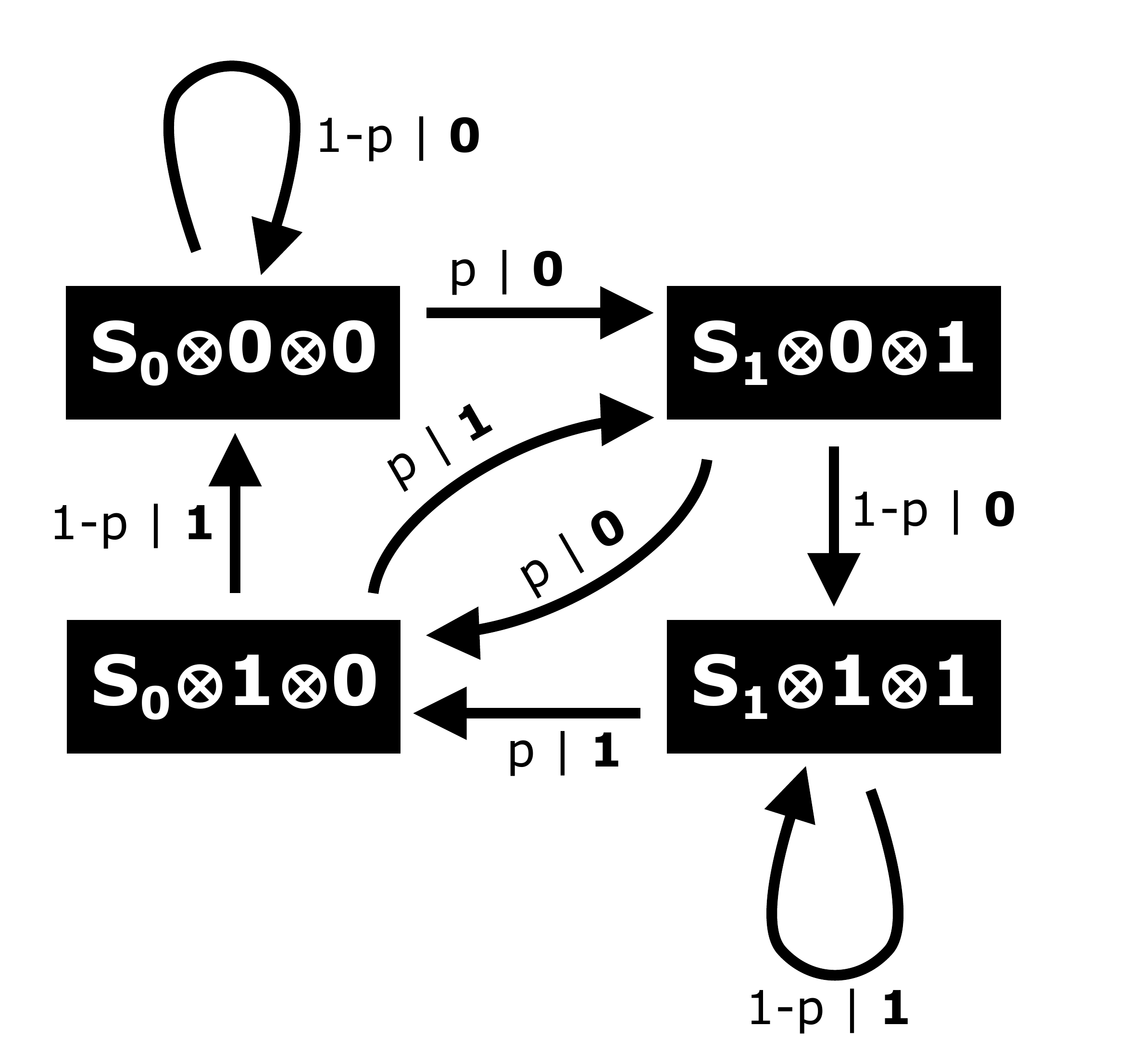}
\subcaption{
\label{fig:PCDBG1}
$K=2$ delay buffer generator.}
\end{subfigure}
} \\
\setcounter{subfigure}{1}
\begin{subfigure}[t]{0.3\textwidth}
\includegraphics[width=\textwidth]{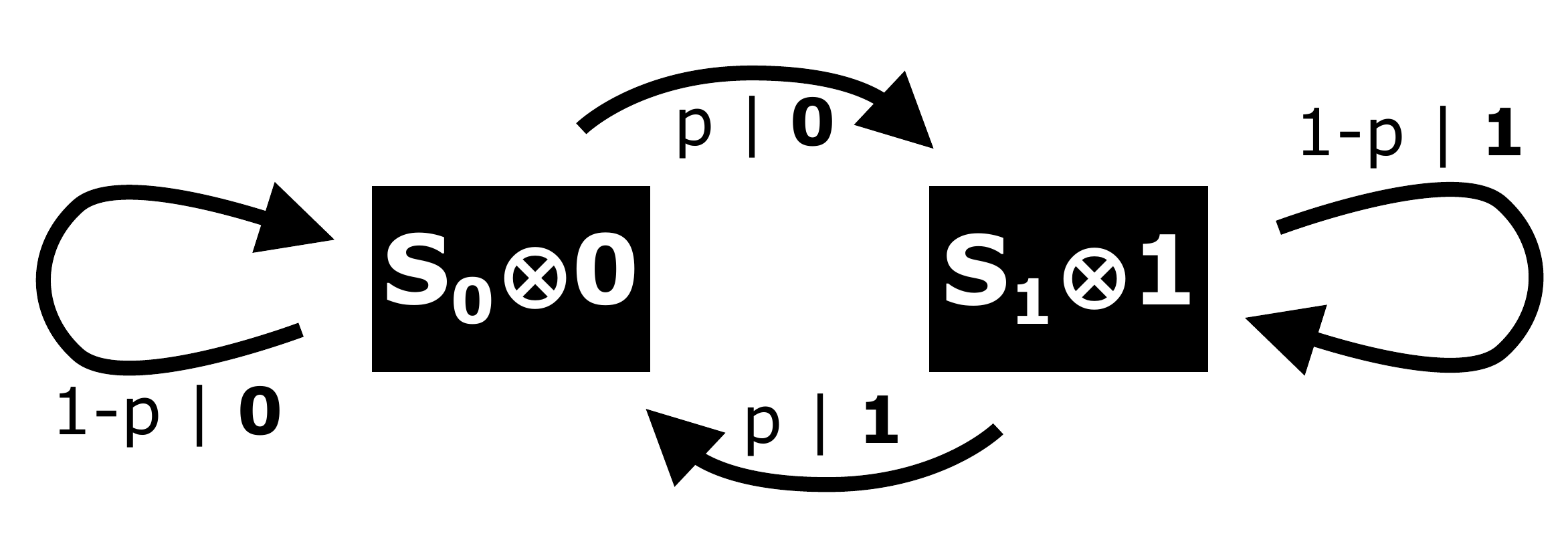}
\subcaption{
\label{fig:PCDBG1}
$K=1$ delay buffer generator. 
}
\end{subfigure}&
\end{tabular}
\end{centering}
\caption{
\label{fig:DB_PC}
\caphead{Delay buffer generators of the perturbed coin.}
The perturbed coin is a binary process generated by repeatedly shaking a fair coin on a plate so that it flips with probability $p$, and then recording whether the coin lands heads-up ({\bf 0}) or tails-up ({\bf 1}).
Its $\varepsilon$--machine (a) and DBGs (b) and (c) can be represented as directed graphs, where each node represents a state of memory, and each directed edge labelled by ``$P \;|\; {\bf a}$'' a transition between the connected states with probability $P$ accompanied by the pattern output ${\bf a}$. 
Such diagrams can be derived for DBGs of arbitrary processes via the algorithm in \cref{sec:DBalgo}.
}
\end{figure*}

Here, we relax this restriction against oracular information, and systematically produce an optimal generator for any process with a finite number of causal states: the {\bf $K$-step delay--buffer generator} (DBG).
Let the alphabet of a pattern $\pastfuture{Y}$ be $\mathcal{Y}$, and of its causal states be $\mathcal{S}$.
The $K$-step DBG has memory $\mathcal{R}$ with the structure $\mathcal{R} :=\mathcal{S}\otimes  \mathcal{Y}^{\otimes K}$ for $K\in\ints^+$,
 such that  $R_{0} := Y_{1} \ldots Y_{K} S_{K}$. 
That is, the memory $R_0$ is composed of a causal state $S_K$ augmented by a {\em delay buffer} of $K$ steps of the pattern $Y_{1:K}$ that immediately precede $S_K$.

\enlargethispage{\baselineskip}
Intuitively, the DBG uses the causal state information within its memory to generate the pattern (e.g.\ by way of a systematically-constructible $\varepsilon$-machine~\cite{CrutchfieldY89}).
However, instead of directly reconfiguring the tape according to the $\varepsilon$-machine's output,
 the DBG stores the $\varepsilon$-machine's output in an internal delay buffer ($\mathcal{Y}^{\otimes K}$).
The entries in this buffer are then cyclically shifted, with the oldest values being emitted as the DBG's output.
This means that the DBG's internal $\varepsilon$-machine operates $K$ steps ahead of the DBG's visible output.
A mechanism by which such memory functions as a generator is detailed in \cref{app:DBM},
 and an example $\varepsilon$--machine and its first two DBGs are drawn as \cref{fig:DB_PC}.
  
\begin{theorem}
Over a free choice of generator memory, the minimum cost $W_g$ of generating $k$ steps of a pattern $\past{Y}$ is given:
\begin{align}
W_g = k \beta W_{tape} = k \left[\Ent{X_{\rm dflt}} - h_Y \right],
\end{align}
where $h_Y$ is the entropy rate of $\pastfuture{Y}$.
\begin{proof}
In \cref{app:DelayBufGenDiss,app:InfiniteCryptic}, I show that the DBG (for large enough delay $K$) has either exactly zero or arbitrarily small $W^{k}_{\rm mem}$.
The claim then follows as a corollary of Lemma 1 and \cref{thm:CrypticDissipation}, since $W^{k}_{\rm mem}=0$ is also the arithmetic minimum a bipartite conditional mutual information can take.
\end{proof}
\end{theorem}

\begin{figure}[bth]
\vspace*{-1em}
\includegraphics[width=0.5\textwidth]{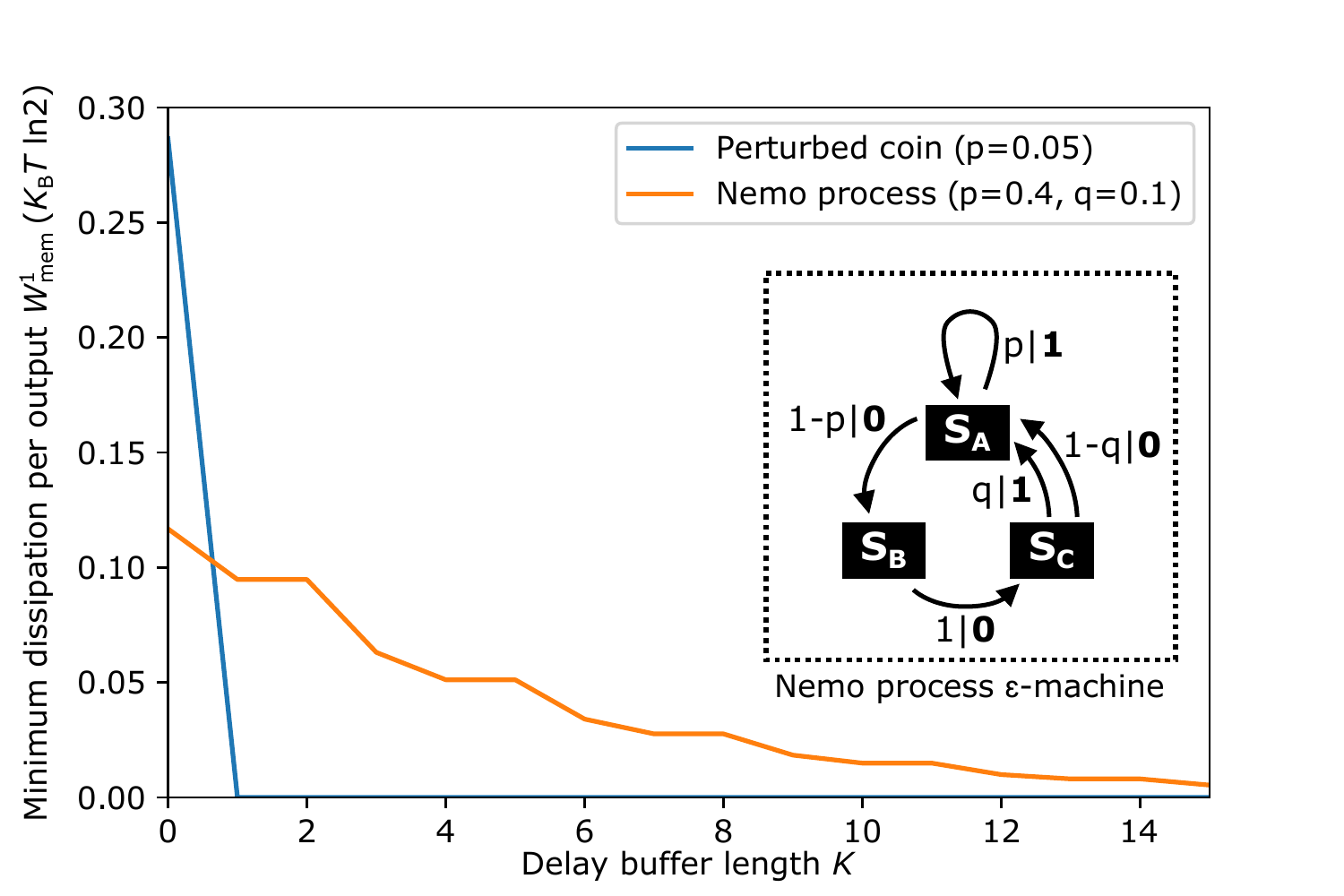}
\caption{
\label{fig:DissipationChart}
\caphead{Minimum dissipation as a function of delay.}
The single--output memory update cost $W^{1}_{\rm mem}$ is plotted for delay buffers generators of different lengths $K$ for the {\em perturbed coin} (see \cref{fig:DB_PC}), and the {\em Nemo process}~\cite{MahoneyEJC11} (see inset $\varepsilon$-machine).
As the perturbed coin has cryptic order $1$, only its $\varepsilon$--machine must dissipate work -- all its non-zero-length DBGs have $W^{1}_{\rm mem}=0$.
The Nemo process has infinite cryptic order, so all its finite-length DBGs are dissipative, but the minimum dissipation decreases monotonically towards zero as the length of the delay buffer increases.
}
\vspace*{-1em}
\end{figure}

\enlargethispage{\baselineskip}
While this thermodynamically efficient generation can also be achieved by other constructions (such as building a generator from the states of a time--reversed $\varepsilon$--machine~\cite{BoydMC18}), 
 this particular construction gives a mechanistic intuition about why such a generator is optimal while the $\varepsilon$--machine (a subcomponent of the DBG) is not.
In particular, the delay length at which the dissipation becomes zero corresponds exactly to the so-called {\em cryptic order}~\cite{MahoneyEJC11} of the pattern (see also \cref{def:cryptic} in \cref{app:DelayBufGen}).
A $K$-DBG with $K$ greater than the cryptic order supplements the causal state with enough extra information to make the machine perfectly {\em retrodictive} and hence avoid the modularity penalty~\cite{BoydMC18}.
The effect of delay length on minimum dissipation is plotted for two example processes in \cref{fig:DissipationChart}\footnote{The source code that generates the graph in \cref{fig:DissipationChart} is available from github via the URL \url{https://github.com/ajpgarner/delay-buffer-generators}.}.

Particularly, the DBG avoids the crypticity--related costs of Theorem~\ref{thm:CrypticDissipation} by updating with the assistance of the previous $K$ steps of the pattern, that (due to \cref{def:FM}) would otherwise be inaccessible to the raw $\varepsilon$--machine.
Cryptic information in a $\varepsilon$--machine corresponds to information recorded about the past that {\em might} be manifest at some point later in the future (and so is essential for statistically accurate generation), 
 but the importance of this information is conditional upon particular sequences being generated at earlier stages of the future.
By retaining up to the cryptic order in a buffer, the DBG can reversibly ``clean up'' this information in its internal $\varepsilon$--machine after the point (namely: the cryptic order) where its clear the information is no longer relevant.

\subsection{Consumers, closed cycles and the second law.}
\label{sec:Consumer}
Let us turn our attention to the consumer of pattern $\pastfuture{X}$ (dashed region C of \cref{fig:Carnot}).
It was shown in \citet{GarnerTVG17} that when the consumer's internal memory $R_0$ corresponds to the causal states of the pattern (i.e. is a realization of the pattern's $\varepsilon$--machine),
 the maximum work $W_c$ that can be extracted by consuming $k$ steps of a pattern with entropy rate $h_Y$, and setting the tape into i.i.d.\ states $X_{\rm dflt}$ is given:
\begin{align}
\beta W_c =  k\left[\Ent{X_{\rm dflt}} - h_Y\right] = k W_{\rm tape}.
\end{align}
Meanwhile, \citet{BoydMC18} show that if a consumer fails to model its input pattern, it will perform worse than this.
Let us complete the argument in the framework of finite models, 
 showing that any hypothetical consumer that exceeds this bound will violate the second law of thermodynamics.

\begin{figure}[hbt]
\vspace{-3em}
\includegraphics[width=0.33\textwidth]{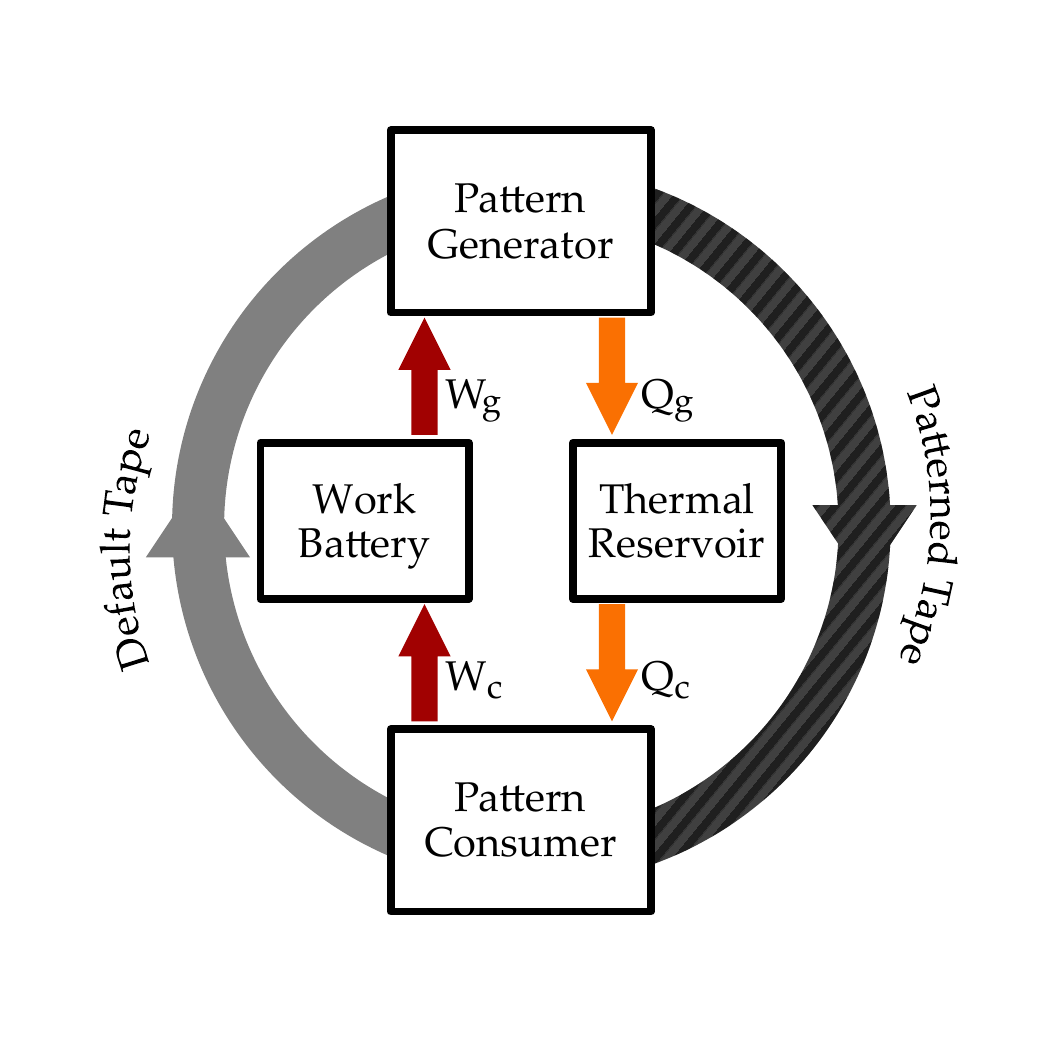}
\vspace{-1em}
\caption{
\label{fig:Carnot2}
\caphead{Pattern cycle with one heat bath.}
This is a specialization of \cref{fig:Carnot} to the case when the generator and consumer operate at the same temperature.
}
\end{figure}

\begin{theorem}
\label{thm:optimal}
The work $W$ that can be extracted by a consumer that reconfigures $k$ steps of pattern $\pastfuture{X}$ with entropy rate $h_X$ into an i.i.d.\ sequence $Y_{\rm dflt}^{\otimes k}$ 
is bounded:
\begin{equation}
W \leq k\left[\Ent{Y_{\rm dflt}} - h_{X}\right].
\end{equation}
\begin{proof}
Consider the closed cycle in \cref{fig:Carnot2} of a generator and consumer of the same pattern $\pastfuture{X}$  (with the same default i.i.d.\ state $\Ent{Y_{\rm dflt}}$),
 both connected to a heat bath at the same temperature $T$. 
Writing $W_{\rm tape} = k\left[\Ent{Y_{\rm dflt}} - h_{X}\right]$,
 from Lemma 1, the generator's work bound is $W_g := W_{\rm tape} + W^k_{\rm mem}$.
Take the assumption (made at the end of \cref{sec:DisGen}) that this bound can be realized.
Meanwhile, let the consumer extract exactly $W_c := k W_{\rm tape} + W_{\rm bonus}$ of work, where $W_{\rm bonus} > 0$ strictly.
The total work exchange in the cycle is thus:
\begin{align}
W_c - W_g & = W_{\rm tape} + W_{\rm bonus} - W_{\rm tape} - W^k_{\rm mem} \nonumber \\
& = W_{\rm bonus} - W^k_{\rm mem} \leq 0
\end{align}
where the last inequality is the Kelvin-Planck statement of the second law.
Hence, to be consistent with the second law, $W_{\rm bonus} \leq W^k_{\rm mem}$ over {\em any} choice of generator memory.

First take the case where $\past{X}$ is a pattern with a finite cryptic order.
Here, a DBG exists with $W^k_{\rm mem} = 0$, so $W_{\rm bonus} \leq 0$, which is a contradiction.
Take the remaining case, where $\past{X}$ has infinite cryptic order.
Here (from Lemma~\cref{lem:ArbitraryLowDissipation} in \cref{app:InfiniteCryptic}), for any $W_{\rm bonus} > 0$ we can always choose a long enough delay such that generator's dissipation $W^{k}_{\rm mem} < W_{\rm bonus}$,
 again giving a contradiction.
Hence, $W_{\rm bonus} \leq 0$, yielding the bound in the claim.
\end{proof}
\end{theorem}

A corollary of this is that it demonstrates the unphysicality of consumers that ``cheat'' by using additional oracular information about their inputs.
\begin{corollary*}
Any consumer that leverages all possible information about the past of the pattern violates the second law of thermodynamics if it is also provided with oracular information about future inputs.
\begin{proof}
The consumer transforms the tape from states $X_{1:k}$ to ${Y_{\rm dflt}}^{\otimes k}$ and updates its memory from $R_0$ to $R_{k}$, effecting the total change in entropy of the tape section and consumer:
\begin{align}
\label{eq:EntCons}
\Delta H = \Ent{R_{k} {Y_{\rm dflt}}^{\otimes k}}  - \Ent{R_0 X_{1:k}}.
\end{align}
The first term of \cref{eq:EntCons} expands to $\Ent{R_{k} {Y_{\rm dflt}}^{\otimes k}} = \Ent{R_{0}} + k \Ent{Y_{\rm dflt}}$ since all systems involved are independent and by stationarity $\Ent{R_k}=\Ent{R_0}$.
The last term expands as $\Ent{R_0} + \cEnt{X_{1:k}}{R_0}$.
When the consumer is a model of the pattern (such that $\cEnt{X_{1:k}}{\past{X}R_0} = \cEnt{X_{1:k}}{R_0}$), but possibly has oracular information, this further expands to $\Ent{R_{0} X_{1:k}} = \Ent{R_0} + kh - \zeta_R\!\left(k\right)$.
The total change in entropy is hence
\begin{align}
\label{eq:ExtractEntropy}
\Delta H & = k\left[\Ent{X_{\rm dflt}} - h\right] + \zeta_R\!\left(k\right) \nonumber \\
& =  k W_{\rm tape}  + \zeta_R\!\left(k\right).
\end{align}
$ \zeta_R\!\left(k\right)$ is non-negative, and thus from \cref{thm:optimal},  $\zeta_R\!\left(k\right)=0$, or the second law is violated.
\end{proof}
\end{corollary*}

\subsection{The thermodynamics of transduction}
\label{sec:Transducer}
By a similar argument to \cref{thm:optimal}, we can quickly derive a loose bound on the work cost of general finite models effecting {\em any} transduction:
\begin{theorem}
The minimum work cost of any finite model that take $k$ steps of a pattern from $X_{1:k}$ to $Y_{1:k}$ is bounded from below by:
\begin{align}
\beta W_{\rm trans} = k \left( h_{X}  - h_{Y} \right)
\end{align}
where $h_{X}$ and $h_{Y}$ are the entropy rates of $\pastfuture{X}$ and $\pastfuture{Y}$ respectively.
\begin{proof}
Consider a cycle of three finite models acting in series on a loop of tape, all connected to the same thermal reservoir:
First, a delay-buffer generator transforming $Z_{\rm dflt}^{\otimes k} \to X_{1:k}$ at cost $W_{g} = \frac{1}{\beta} k \left[\Ent{Z_{\rm dflt}} - h_X \right]$;
second, an arbitrary transducer transforming $X_{1:k}$ to $Y_{1:k}$ at cost $W$;
third, an $\varepsilon$-machine consumer transforming $Y_{1:K}$ to $Z_{\rm dflt}^{\otimes k}$ recovering work $W_{c} =  \frac{1}{\beta}  k \left[\Ent{Z_{\rm dflt}} - h_Y\right]$.
The existence of the first and third machines are guaranteed for any $\pastfuture{X}$, $\pastfuture{Y}$ by construction 
 -- and moreover both effect the desired transformation in the thermodynamically optimal manner.
Applying the Kelvin-Planck statement of the second law to this cycle:
\begin{align}
k \left[\Ent{Z_{\rm dflt}} - h_X \right]  + W - k \left[\Ent{Z_{\rm dflt}} - h_Y\right] & \geq 0,
\end{align}
and hence:
$W  \geq  k \left(h_X - h_Y\right)$.
\end{proof}
\end{theorem}

This bound holds over all choices of memory, but is unlikely to be tight for any particular memory choice (e.g.\ \cref{thm:CrypticDissipation} provides examples where it is not tight for certain generators).
Moreover, it is not here proven whether this bound is tight even for {\em all} choices of transduction (i.e.\ whether a perfect choice of memory can always be found).
However, we remark that (via \mbox{Eq.~(5)} of \citet{BoydMC16}) this bound can be saturated by any transduction that can be implemented as an {\em information ratchet}
  -- a particular type of finite model that effects the manipulation by way of thermal fluctuations on the coupled tape--memory system.
This identifies such devices as thermodynamically optimal where they exist.

With these results, we establish the setting of finite models interacting via tapes as an well-defined {\em information reservoir} framework.
Each section of tape can be thought of as a special type of battery (the information reservoir) that can be perfectly discharged only by an agent with the appropriate memory.
Since the framework {\em only} permits models to communicate via the tape (and not, e.g.,\ allowing one model to directly access the internal memory of another),
 the work cost of the action of any finite model can then be split into two components:
 \begin{enumerate}
\item A ``reversible'' work cost `$k \left(h_Y - h_X\right)$'' that is a function of the change in tape entropy rate that can {\em always} be recovered by another model (e.g.\ the $\varepsilon$-transducer consumer) that subsequently receives the tape.
Since this is freely interchangeable with work stored in a battery, we identify this portion of the work cost as that spent to alter the {\em free energy} of the tape.
\item A non-negative excess work cost (e.g. $W^{k}_{\rm mem}$ in the generator) that is a function of the particular choice of memory of the finite model, which can {\em never} be recovered by any thermodynamically consistent finite model.
This can thus be regarded as a type of {\em dissipated work}.
\end{enumerate} 

\section{Discussion}
\subsection{The thermodynamics of forecasting.}
\label{sec:Forecast}
Recall Theorem~\ref{thm:CrypticDissipation}: the excess work cost of generation is bounded by the {\em discarded cryptic information}: information remembered by the generator about the pattern's history, which never manifests in the pattern's future, and was subsequently expunged from memory.
This is conceptually similar to the unavoidable dissipation caused by ``useless nostalgia'' presented by \citet{StillSBC12}.
Indeed, in certain limits, the results here and of Still {\em et al.}\ describe the same physical phenomenon.

Still et al.\ consider a (bottom-up) setting motivated by fluctuation theorem literature~\cite{Jarzynski97,Crooks99}, in which a system is driven between its internal states by an external signal.
In the finite model framework, the role of internal states can be played by model memory,
 and the role of the external signal by a pattern.
It is insufficient to {\em only} consider the input--output behaviour of this system, since it does nothing to the pattern (admitting a trivial, memoryless transducer).
Instead, we consider a new type of finite model that captures both the driven dynamical behaviour {\em and} the capacity to predict:
\begin{definition}
A {\bf forecaster}\footnote{
I deviate from the canonical word ``predictive model'' for this {\em specific} do-nothing machine to stress the difference in its operational behaviour between the (destructive, perfect) consumer, 
 and (non-oracular) generators -- all of which could be called predictive models within computational mechanics literature.
}
 of pattern $\pastfuture{X}$ is a finite model that reads the pattern $k$ steps at a time without altering it,
 in such a way that the model's internal memory $R$ can be used (at any time) to initialize a statistically--accurate generator of $\future{X}$ (i.e.\ satisfying $\cPr{\future{X}}{R} = \cPr{\future{X}}{\past{X}}$).
\end{definition}

A forecaster is neither strictly a generalization or a specialization of the driven system in \citet{StillSBC12},
 though there is an intersection between the schemes.
A forecaster does not make mechanistic assumption as to what constitutes heat exchange or work exchange (i.e.\ taking a top-down approach),
 nor is a forecaster restricted to moving forward step of the pattern at a time.
Conversely, we {\em do} make the additional restriction that a forecaster has perfectly predictive memory (i.e.\ capturing all of $\Info{\past{X}}{\future{X}}$).

\citet{StillSBC12} calculate the following quantities to bound the work cost associated with the signal advancing from $X_0$ to $X_1$:
\begin{align}
I_{\rm mem} & := \Info{R_0}{X_0}, \\
I_{\rm pred} & := \Info{R_0}{X_1}, \\
\beta W_{\rm diss} & = I_{\rm mem} - I_{\rm pred}.
\label{eq:StillResult}
\end{align}
The RHS of this last equation is referred to by the authors as ``useless instantaneous nostalgia'',
 because it represents the difference between the information that the driven system remembers about the previous symbol ($I_{\rm mem}$) and that it has about the next ($I_{\rm pred}$).

Let us compare this quantity to the entropy change of the forecaster calculated in this article's framework:
\begin{align}
\Delta H & = \Ent{R_k X_{1:k}} - \Ent{R_0 X_{1:k}} \nonumber \\
& = \Info{R_0}{X_{1:k}} - \Info{R_k}{X_{1:k}} \nonumber \\
& = \Info{R_0}{X_{1:k}} - \Info{R_0}{X_{k-1:0}}
\label{eq:ForecastEntropy}
\end{align}
where the first step is an expansion of the definition of mutual information, and the second follows from stationarity.
Recalling that $W \propto -\Delta H$, and noting that, as the entropy rate of the tape is unchanged {\em all} work done here can considered as a dissipation (see \cref{sec:Transducer}) 
 we see that this gives the exact same bound as \cref{eq:StillResult} when $k=1$.

Finally, we can rewrite this expression in form similar to \cref{thm:CrypticDissipation}:
\begin{align}
\label{eq:EntropyForecast}
\Delta H & = - \cInfo{\past{X}}{R_0}{\future{X}R_{k}} + \zeta_R(k).
\end{align}
(Proof in \cref{app:Forecast}.)
Here, the first term is exactly the same {\em discarded cryptic information}, as was responsible for the fundamental lower bound on dissipation during generation (per Theorem~\ref{thm:CrypticDissipation}) 
 -- but there is also the term $\zeta_R(k)$ -- the {\em oracular information} that the forecaster holds about the upcoming word of the pattern.
In \mbox{Still et al.'s} setting $\zeta_R(k)=0$ by construction, and the two expressions are the same.

\subsection{Conclusion and outlook}
In this article, we examined the thermodynamic consequences of manipulating patterns with finite models.
We saw that it is the {\em discarding of cryptic information} -- stored information about a pattern's history that never shows up in its future behaviour -- that is responsible for heat dissipation in pattern generators.
We also considered a systematic construction that could avoid this cost: the delay buffer generator,
 which internally produces the pattern ahead of time, but delays its output until it is sure it no longer needs the information to reversibly update its internal state.
The minimum length of such delay corresponded to the {\em cryptic order}~\cite{MahoneyEJC11}  of the pattern, imbuing a hitherto highly information--theoretical term with additional physical meaning.
Finally, we touched on the role of consumers, and other more general devices in a framework of finite models communicating via their tapes,
 and showed how this formed an ``information reservoir'' framework, where the cost of running each finite model could be split into two portions:
 a recoverable ``free energy'' associated changing the tape's entropy rate, and an irreversible dissipation arising from the specific choice of memory.
 
These results suggest several directions for future research.
First is to produce the tight memory--specific bounds on consumers in the manner of \cref{thm:CrypticDissipation} (taking into account the different set of applicable data processing inequalities).
These will likely be a function of the model's failure to predict its input~\cite{BoydMC17_b,BoydMC18}.
Likewise, one might search for the general dissipative cost of {\em any} finite model -- though the corollary of \cref{thm:optimal} shows the need to be careful about what memory choices are permissible 
 (suggesting perhaps a need to restrict to finite models whose action may be described by a non-antipicatory transducer~\cite{BarnettC15}).
Finally, the computational mechanical framework can be extended to quantum systems (e.g.\ \cite{GuWRV12,MahoneyAC16,SuenTGVG17,BinderTG18}),
 and recent effort considers thermodynamic costs in this light~\cite{LoomisC20}.
This invites the natural question: how do quantum input--output devices~\cite{ThompsonGVG17} -- a quantum analogue to finite models --  thermodynamically perform?

\acknowledgments
I am grateful for discussions with Felix~Binder, Alec~Boyd, Thomas~Elliott, Mile~Gu, Marius~Krumm, Jayne~Thompson, and Paul~Riechers.
This project was made possible through the support of a grant from the John Templeton Foundation. 
The opinions expressed in this publication are those of the author and do not necessarily reflect the views of the John Templeton Foundation.
This research was supported through the grants FQXi-RFP-1815 ``Where agents and algorithms meet...''\ and FQXi-RFP-IPW-1903 ``Are quantum agents more energetically efficient at making predictions?''\ from the Foundational Questions Institute and Fetzer Franklin Fund, a donor advised fund of Silicon Valley Community Foundation, as well as the National Research Foundation (NRF), Singapore, under its NRF Fellow program (Award No.\ NRF-NRFF2016-02).

%
\balancecolsandclearpage


\appendix
\section{Information anatomy of finite models}
\label{app:InfoAnatomyWhole}
\subsection{Information diagrams (brief overview).}
\label{app:InfoDiagrams}
In the following sections, we will be considering the relationships between many random variables.
Writing $A_i$, $B_i$ to denote arbitrary random variables,
 the pertinent set of information quantities are the {\em multivariate mutual information}: 
 \begin{align}
\label{eq:mvInfo}
\iInfo{A_1}{\ldots}{A_N}  \quad & := \hspace{1em} - \hspace{-2em}\sum_{C\subseteq \{A_1,\ldots A_N\}} \hspace{-1.5em} (-1)^{|C|} \; \Ent{C},
\end{align}
(where the sum is over the power set of $\{A_1, \ldots, A_N\}$)
and the {\em conditional multivariate mutual information}:
\begin{align}
\label{eq:cmvInfo}
\mvcInfo{A_1}{\ldots}{A_N}{B_{1} \ldots B_{M}}  & :=  \nonumber\\
& \hspace{-8em} - \hspace{-2em}\sum_{C\subseteq \{A_1,\ldots A_N\}} \hspace{-1.5em} (-1)^{|C|}\; \cEnt{C}{B_{1}\ldots B_M}.
\end{align}

The cases $N=1$ (and $M=1$) corresponds to usual definition of (conditional) entropy.
Likewise, $N=2$ (and $M=1$) is the traditional (conditional) bipartite mutual information.
However, when $N>2$, the information quantities can be positive, negative or zero.

There is an isomorphism between these measures and the distinct regions of a Venn diagram whose primary sets (i.e.\ largest ``rings'') represent random variables~\cite{Hu62}.
Particularly: the region in the intersection of $A_1$\ldots $A_N$ corresponds to $\iInfo{A_1}{\ldots}{A_N}$,
 and the portion of that intersection that excludes the union of $B_1 \ldots B_M$ corresponds to $\mvcInfo{A_1}{\ldots}{A_N}{B_{1} \ldots B_{M}}$.
Meanwhile, the union of sets $A_1\ldots A_k$ corresponds to the joint entropy $\Ent{A_1\ldots A_k}$.
This allows for the rapid derivation of equalities between the various information quantities in a visual manner: 
 the information quantity associated with any region must correspond to sum of the information quantities associated with its constituent parts.

\subsection{Classes of information in finite models}
\label{app:Classes}
With respect to a pattern $\pastfuture{Z}$, the information in memory $R$ can be divided into four {\em pattern--memory} classes~\cite{CrutchfieldEJM10,EllisonMJCR11} (\cref{fig:VennSimple}):-
\vspace*{-0.5em}
\begin{itemize}
\item {\em Predictive information}: $E_R := \iInfo{\past{Z}}{\future{Z}}{R}$; information from the pattern's past, stored in the memory,
 about the pattern's future.
\item {\em Cryptic information}: $\chi_R := \cInfo{\past{Z}}{R}{\future{Z}}$; information from the pattern's past, stored in the memory, but never manifest in the future.
\item {\em Oracular information}: $\zeta_R := \cInfo{\future{Z}}{R}{\past{Z}}$; information about the pattern's future stored in the memory, but not predictable from the pattern's past.
\item {\em Gauge information}: $\varphi_R := \cEnt{R}{\past{Z},\future{Z}}$; information in the memory that has nothing to do with the pattern.
\end{itemize}

When $R$ is a generator of $\pastfuture{Z}$, then $\cInfo{\past{Z}}{\future{Z}}{R} = 0$. 
Then, $E_R = \iInfo{\past{Z}}{\future{Z}}{R} = \Info{\past{Z}}{\future{Z}} =: \mathcal{E}$, the so-called {\em excess entropy} of the pattern, 
 and hence $E_R \geq 0$.
$\chi_R$, $\zeta_R$ and $\varphi_R$ are also always non-negative.
Thus, $\Ent{R} = E_R + \chi_R + \eta_R + \varphi_R$.

\begin{figure}[htb]
\centering
\includegraphics[width=0.35\textwidth]{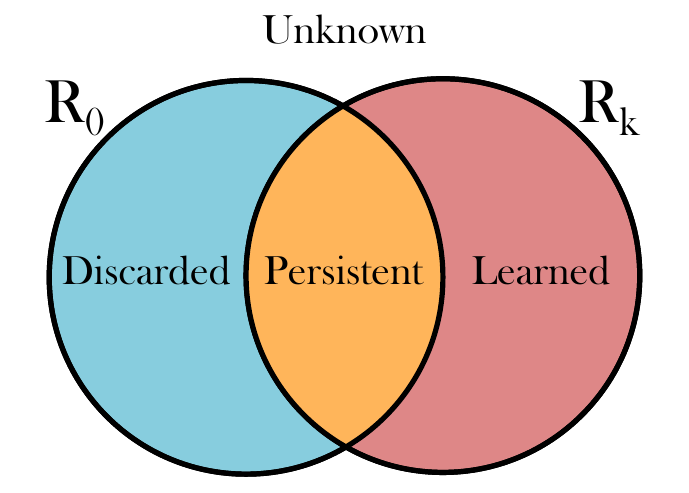}
\caption{
\label{fig:MUC}
\caphead{Memory--update classes.}
A classification of the change in information between times $0$ and $k$.
}
\end{figure}

Meanwhile, when the memory updates from time $0$ to $k\in\ints^+$, 
 there is another implied information diagram of four mutually--exclusive {\em memory--update classes} (\cref{fig:MUC}):-
\begin{itemize}
\item {\em Unknown information} that is not involved with the memory at either time, 
\item {\em Learned information} present at time $k$ but not at time $0$, 
\item {\em Discarded information} present at time $0$ but not at time $k$,
\item {\em Persistent information} present at both times.
\end{itemize}

\subsection{Information anatomy of a generator}
\label{app:Anatomy}
\begin{figure}[htb]
\centering
\includegraphics[width=0.45\textwidth]{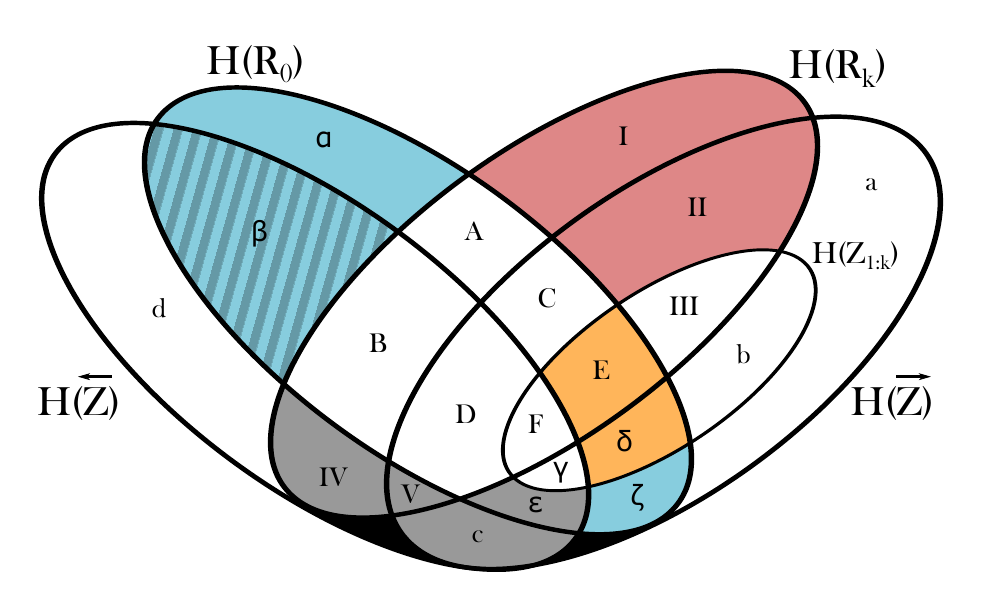}
\caption{
\label{fig:MemoryUpdate}
\caphead{Generator memory update.}
See classification in \cref{app:Anatomy}.
The minimum dissipation of a generator is proportional to the striped blue region $\beta$.
The red, blue and yellow regions respectively correspond to 
 $\cEnt{R_{k}}{Z_{1:k} R_0}$, $\cEnt{R_{0}}{Z_{1:k} R_{k}}$ and $\zeta_R\!\left(k\right) =$ $ \cInfo{Z_{1:k}}{R_0}{\past{Z}}$  (as in \cref{eq:GenDiss}).
For a generator, the gray regions are always $0$ (see \cref{app:Zero}).
}
\end{figure}

By considering the overlap between $\past{Z}$, $\future{Z}$, $Z_{1:k}\subset \future{Z}$, $R_0$ and $R_k$,
 we arrive at figure \cref{fig:MemoryUpdate}.
In the following, we interpret the regions of this figure in terms of the classes outlined in \cref{app:Classes}.
 
First, we list the regions of {\bf unknown information} (lowercase labels), not involved with the model memory at either time:
\begin{enumerate}[label={\em \alph* } -- ]
\item the {\em unknowable future} -- the randomness in the pattern that cannot be predicted either from the past, or from the memory at either time step (this region will generally be infinite in magnitude).
\item the randomness in $Z_{1:k}$ that could not be predicted from the past, and was also not predictable from the memory at time $0$, and was also not stored in the new state of the memory at time $k$. (When $R$ is a model of $\pastfuture{Z}$, this will coincide with the {\em ephemeral information} of \citet{JamesEC11}).
\item the information $\past{Z}$ contains about $\future{Z}$ that was not stored in the memory at either time $0$ or $k$. 
\item the {\em forgotten, irrelevant past} -- information about $\past{Z}$ that has no relation to any part of the future, and is not in the memory at either time $0$ or $k$ (this region will generally be infinite in magnitude).
\end{enumerate}
Additionally, the unlabelled space around the diagram fits trivially into this memory--update class.

Next, we list the {\bf learned information} (uppercase Roman numerals) not present in $R_{0}$, but present in $R_{k}$:
\begin{enumerate}[label={\em \Roman* } -- ]
\item new {\em gauge information}, which does not relate to any part of the pattern, past or future.
\item new {\em oracular information}, pertaining to parts of $\future{Z}$ that is not visible through any part of the pattern up to and including step ${k}$.
\item new information learned about the word of the pattern $Z_{1:k}$ just manipulated.
(Some of this may subsequently become cryptic with respect to memory time $k$, and some becomes predictive -- this distinction is not visible in this diagram). 
\item the information about the past $\past{Z}$ not manifest in $R_0$ that suddenly becomes visible in $R_{k}$. 
\item new information that the past contained about the future. 
\end{enumerate}

Now, we enumerate the {\bf discarded information} (Greek labels) present in $R_0$, but not present in $R_{k}$:
\begin{enumerate}
\item[{\em $\alpha$ --}]  {\em discarded gauge information}, which does not relate to any part of the pattern, past or future.
\item[{\em $\beta$ --}] {\em discarded cryptic information}, not related to any part of the future of the pattern, but that is related to the past. 
This quantity governs the minimum dissipation for generators (\cref{thm:CrypticDissipation}).
\item[{\em $\gamma$ --}] {\em used and discarded predictive information}, which was visible from the past, used in the generation of $Z_{1:k}$, but not carried forward in the memory at time $k$.
\item[{\em $\delta$ --}] {\em used and discarded oracular information}, which was not visible from the past, but was used in the generation of $Z_{1:k}$ and not carried forward in the memory at time $k$.
\item[{\em $\epsilon$ --}] {\em wasted predictive information}, pertinent to parts of the future from times $k+1$ onwards, but discarded before it has been used to act on these parts of the pattern.
\item[{\em $\zeta$ --}] {\em wasted oracular information}, pertinent to parts of the future from times $k+1$ onwards that was stored in $R_0$ and not otherwise visible from the past -- but that was discarded before it could be used (because it has not been transmitted to $R_{k}$). 
Although obviously wasteful, there is no reason to rule this out {\em a priori}.
\end{enumerate}

Finally, we list {\bf persistent information} (uppercase labels) present in the memory at both times $0$ and $k$:
\begin{enumerate}[label={\em \Alph* } -- ]
\item {\em persistent gauge information}, which does not relate to any part of the pattern, past or future. 
If one views $R$ as a hard disk, and the update mechanism as changing one file on that disk relating to the pattern $\pastfuture{Z}$; this region would be all the other unrelated files on that disk.
\item {\em persistent cryptic information}, related to the past of the pattern, but unrelated to the future.
\item {\em persistent oracular information}, related to the future of the pattern, but not visible from either the past or the newly output word $Z_{1:k}$.
\item {\em persistent predictive information}, related to the future of the pattern, and visible from the past, but not related to the most recent word $Z_{1:k}$.
\item {\em used and stored oracular information}. This is the information that was oracular at time $t$; but has since become visible in the most-recently manipulated word $Z_{1:k}$, such that at time $k$ it is no longer oracular. 
At this point, it will either have become part of the predictive information (if it relates to $Z_{k+1}$ onward), or otherwise become cryptic. This distinction is not shown on the diagram\footnote{This would require also illustrating the set $\future{Z}_k := \lim_{L\to\infty} Z_{k+1:K+L}$}.
\item {\em used and stored predictive information}. This is information visible from the history up to time $0$, and used in the recently manipulated $Z_{1:k}$. 
At time $k$, some of this information may become purely cryptic (i.e.\ unrelated to $Z_{k+1}$ onward), whereas some may still be relevant to the future (and remain predictive). 
This distinction is not shown on the diagram.
\end{enumerate}

\section{Data processing and generators}
\label{app:DPIZeros}
\subsection{The data-processing inequality}
\label{app:DPI}
Three random variables $X$, $Y$ and $Z$ (over respective alphabets $\mathcal{X}$,$\mathcal{Y}$,$\mathcal{Z}$) form a {\em Markov chain} written $X\to Y\to Z$ if for all $x\in\mathcal{X}$, $y\in\mathcal{Y}$, $z\in\mathcal{Z}$:
\begin{align}
\label{eq:Markov}
\Pr{X\!=\!x, Y\!=\!y, Z\!=\!z} & = \nonumber\\
& \hspace{-8em} \Pr{X\!=\!x} \cPr{Y\!=\!y}{X\!=\!x}\cPr{Z\!=\!z}{Y\!=\!y}.
\end{align}
This property is guaranteed if there is a (generally non-deterministic) map $f: \mathcal{Y} \to \mathcal{Z}$ such that for each $y\in\mathcal{Y}$ the outcome $z\in\mathcal{Z}$ is obtained with probability depending only on the state of $Y$\footnote{
The only important constraint is that $\cPr{Z\!=\!z}{Y\!=\!y, X\!=\!x} =\cPr{Z\!=\!z}{Y\!=\!y}$,
  and so it is irrelevant whether or not there is a natural map from $X\to Y$. 
}.
Markov chains are subject to the {\em data--processing inequality} (DPI) (e.g.~\cite{CoverT91} \S{}2.8):
\begin{align}
\label{eq:dpi}
\Info{X}{Y} \geq  \Info{X}{Z},
\end{align}
with equality holding if and only if $\cInfo{X}{Y}{Z} = 0$.

If $X=AD$, $Y=BD$ and $Z=CD$ form the Markov chain $AD\to BD \to CD$,
  substitution into the above gives a similarly useful expression:
\begin{align}
\label{eq:cdpi}
\cInfo{A}{B}{D} \geq \cInfo{A}{C}{D}
\end{align}
after making the expansions $\Info{AD}{BD} = \Ent{D} + \cInfo{A}{B}{D}$ and $\Info{AD}{CD} = \Ent{D} + \cInfo{A}{C}{D}$.

Finally, whenever $X\to Y\to Z$ form a Markov chain, then also~\cite{CoverT91}
\begin{align}
\label{eq:dpe}
\cInfo{X}{Z}{Y} = 0.
\end{align}
We refer to this equation as the {\em data-processing equality} (DPE), since it is intrinsically related to Ineq.~\eqref{eq:dpi}.

\subsection{Empty regions of \cref{fig:MemoryUpdate}}
\label{app:Zero}

\begin{lemma}
\label{lem:zeros_1}
The quantities represented by regions $IV$, $V$, $c$ and $\varepsilon$ of \cref{fig:MemoryUpdate} are all zero.
\begin{proof}
First recall (as discussed in \cref{sec:mem}) that $\past{Z} \to R_0 \to \future{Z}$ is a Markov chain.
Thus, immediately from the DPE  (\cref{eq:dpe}) $\cInfo{\past{Z}}{\future{Z}}{R_0} = 0$,
 which expressed as regions in \cref{fig:MemoryUpdate} is: 
\begin{align}
c + V & = 0.
\end{align} 

Next, note that a finite generator can also produce $Z_{1:k} R_k$ from its initial memory state $R_0$,  so there is a map $f: R_0 \mapsto Z_{1:k} R_k$.
Similarly, there is an implied map $f': R_0 \mapsto R_k$, which can be calculated from $f$ by taking the marginals over $Z_{1:k}$.
Thus, $\past{Z}\to R_0 \to R_k$ is also a Markov chain 
 and the DPE (\cref{eq:dpe}) implies $\cInfo{\past{Z}}{R_k}{R_0} = 0$.
When expressed regions in \cref{fig:MemoryUpdate}:
\begin{align}
IV + V & = 0.
\end{align}

Now, we will show $IV=0$, by considering the two (composite) random variables $T_1 := R_0 R_0 \future{Z} $ and $T_2 := R_k R_0 \future{Z}$.
Though the map $h: T_1 \to T_2$ is not directly the action of the generator, 
 we can still use the generator's action to guarantee the mathematical existence of a map, such that $X\to T_1 \to T_2$ (for arbitrary $X$) is a Markov chain.
Particularly, we have the maps $f: R_0 \to Z_{1:k} R_k$, and $g:R_0 \to \future{Z}$ from the generator,
 such that the first $k$ steps in the latter agree statistically with the term $Z_{1:k}$ in the former.
Thus, we may simply extend the map $f$ to $h'=f\otimes\id\otimes\id$ taking $R_0 R_0 \future{Z}$ to $R_k Z_{1:k} R_0 \future{Z}$,
  and thus (via the trivial map $Z_{1:k}\future{Z}$ to $\future{Z}$) also a map $h: R_0 R_0 \future{Z} \to R_k R_0 \future{Z}$.
Now, specializing $X:= \past{Z} R_0 \future{Z}$,  we may substitute $A=\past{Z}$, $B=R_0$, $C=R_k$ and $D=R_0\future{Z}$ to form Markov chain $\past{Z}R_0\future{Z} \to R_0 R_0 \future{Z} \to R_k R_0 \future{Z}$
 and use Ineq.~\eqref{eq:cdpi} to derive:
\begin{align}
\label{eq:MemoryDPI_V}
0 = \cInfo{\past{Z}}{R_0}{R_0\future{Z}} & \geq \cInfo{\past{Z}}{R_k}{R_0 \future{Z}} = 0.
\end{align} 
The first equality uses $\cInfo{X}{Y}{YZ} = 0$, and the last follows because $\cInfo{\past{Z}}{R_k}{R_0 \future{Z}}$ is non-negative, as a bipartite conditional mutual information.
Expressed as a region in \cref{fig:MemoryUpdate}:
\begin{align}
IV = 0.
\end{align}

Since $IV+V = 0$, $V+c=0$  and $IV=0$, it follows:
\begin{align}
V = 0, &\qquad c =0.
\end{align}

Finally, $\varepsilon$ of \cref{fig:MemoryUpdate} represents the quantity $\mvcInfo{R_0}{\past{Z}}{\future{Z}}{R_k Z_{1:k}}$.
From definition \cref{eq:cmvInfo} this may be re-expressed as  $\varepsilon= \mvcInfo{R_0}{\past{Z}}{\future{Z}_{k+1}}{R_k Z_{1:k}}$
 after writing $\future{Z}_{k+1} := Z_{k+1} Z_{k+2} \ldots$ and noting $\future{Z} = Z_{1:k} \future{Z}_{k+1}$.
We can then write\footnote{This region and its divisions are not visible in \cref{fig:MemoryUpdate}, since the set associated with $\future{Z}_{k+1}$ is not drawn.}
\begin{align}
\label{eq:expandWEps}
\cInfo{R_0}{\future{Z}_{k+1}}{R_k Z_{1:k}} & = \cInfo{R_0}{\future{Z}_{k+1}}{R_k Z_{1:k} \past{Z}} + \varepsilon.
\end{align}
By definition, a generator admits maps $R_0 \to R_k Z_{1:k}$ and $R_k \to \future{Z}_{k+1}$ (the latter from stationarity, and the map $R_0\to \future{Z}$).
Thus, $R_0 \to R_k Z_{1:k} \to \future{Z}_{k+1}$ is a Markov chain, and the LHS of \cref{eq:expandWEps} is $0$ by the DPE (\cref{eq:dpe}).
Meanwhile, we argue for the existence of a map from $R_k\past{Z}Z_{1:k} \ to \future{Z}_{k+1}$.
This is done via stationarity and the map $\past{Z}R_0 \to \future{Z}$,
 which in turn is defined by ignoring $\past{Z}$ completely and generating $\future{Z}$ exclusively from $R_0$ (valid because $\past{Z}\to R_0 \to \future{Z}$ is a Markov chain).
Then,  $R_0 \to R_k \past{Z} Z_{1:k} \to \future{Z}_{k+1}$ is another Markov chain (recall: once we have a fully defined map, the first random variable in the chain is irrelevent),
  and the first quantity on the RHS of \cref{eq:expandWEps} is also $0$ by the DPE.
We hence conclude 
\begin{align}
\varepsilon=0.
\end{align}

\end{proof}
\end{lemma}

\begin{lemma}
The set $Z_{1:k}$ has no intersection with region $c$, or with region $V$ as defined in \cref{fig:MemoryUpdate}.
\begin{proof}

Consider the bisection of region $c = \cInfo{\past{Z}}{\future{Z}}{R_0 R_k}$ with $Z_{1:k}$, which creates two (signed) quantities:
\begin{align}
x & := \cInfo{\past{Z}}{\future{Z}}{R_0 R_k Z_{1:k}} \geq 0 \\
\tilde{x} & := \mvcInfo{\past{Z}}{\future{Z}}{Z_{1:k}}{R_0 R_k} = -x
\end{align}
where the first inequality is true for any conditional bipartite mutual information,
 and the last equality follows from Lemma~\ref{lem:zeros_1} (noting $x+\tilde{x}=c=0$).
Thus, $\tilde{x} \leq 0$.

From the definition in \cref{eq:cmvInfo}:
\begin{align}
\label{eq:elision}
\mvcInfo{A}{BC}{C}{D}  \hspace{-2em} & \hspace{2em} := \cEnt{ABCC}{D} - \cEnt{ABC}{D}  \nonumber \\ 
& \quad-  \cEnt{BCC}{D} - \cEnt{AC}{D} \nonumber \\
& \quad + \cEnt{A}{D} + \cEnt{BC}{D} + \cEnt{C}{D} \nonumber \\
& = -\cEnt{AC}{D} + \cEnt{A}{D} + \cEnt{C}{D} \nonumber \\
& = \cInfo{A}{C}{D} \geq 0.
\end{align}
Then, since $\future{Z} = Z_{1:k} \future{Z}_{k+1}$ (where $\future{Z}_{k+1} = Z_{k+1} Z_{K+2}\ldots$), 
 we substitute in $A=\past{Z}$, $B=\future{Z}_{k+1}$, $C={Z_{1:k}}$ and $D=R_0 R_k$
 to arrive at 
 \begin{align}
 0 \geq \tilde{x} & = \cInfo{\past{Z}}{Z_{1:k}}{R_0 R_k} \geq 0.
\end{align}
Thus, $\tilde{x}=0$,
 and  $Z_{1:k}$ has no intersection with $c$.

Next, we bisect region $V= \mvcInfo{\past{Z}}{\future{Z}}{R_k}{R_0}$ with $Z_{1:k}$,
 again creating two equal and opposite ($V=0$, Lemma~\ref{lem:zeros_1}) quantities:
\begin{align}
 y & := \mvcInfo{\past{Z}}{\future{Z}}{R_k}{R_0 Z_{1:k}}, \\
\tilde{y} & = \insaneInfo{\past{Z}}{\future{Z}}{R_k}{Z_{1:k}}{R_0} = -y.
\end{align}

Rather than addressing this quantity directly, it is easier to consider region $\tilde{x}+\tilde{y}= \mvcInfo{\past{Z}}{\future{Z}}{Z_{1:k}}{R_0}$,
 and once more use \cref{eq:elision} with the substitutions $A = \past{Z}$, $B=\future{Z}_{k+1}$, $C=Z_{1:k}$, $D=R_0$ to write
\begin{align}
\tilde{y} = \tilde{x}+\tilde{y} = \cInfo{\past{Z}}{Z_{1:k}}{R_0},
\end{align}
(where the first equality uses $\tilde{x} = 0$, as just proven).

Since the generator's definition gives the Markov chain $\past{Z}\to R_0 \to Z_{1:k}$,
 we may immediately use the DPE (\cref{eq:dpe}) to set $\tilde{y}=0$ (and hence also $y=0$).
Thus, $Z_{1:k}$ has no intersection with $V$.
\end{proof}
\end{lemma}

\section{Stationary update in generators}
\label{app:ToG}

For any cyclically--operating (i.e.\ stationary) generator of a stationary pattern, the pattern--memory classes of information in the memory should remain constant in time.
Thus, using \cref{fig:MemoryUpdate} to examine the constitution of the memory at times $0$ and $k$, we can identify the equalities summarized in the following lemma:

\begin{lemma}[Conservation of information by class]
\label{lem:cons}
For a stationary process manipulating $\pastfuture{Z}$ using memory $R$:
\begin{enumerate}[label=\roman*.]
\item from conservation of gauge information:
\begin{align}
\label{eq:ConserveGaugeInfo}
\cEnt{R_0}{R_{k} \pastfuture{Z}} & = \cEnt{R_k}{R_0 \pastfuture{Z}},
\end{align}
\item from conservation of oracular information:
\begin{align}
 \cInfo{R_0;R_{k}}{Z_{1:k}}{\past{Z}} + \cInfo{R_0}{Z_{1:k}}{\past{Z} R_{k}} \hspace{-17em} & \nonumber \\
& \qquad + \cInfo{R_0}{\future{Z}}{\past{Z} Z_{1:k} R_{k}}  = \cInfo{R_k}{\future{Z}}{R_0 \past{Z} Z_{1:k}},
\label{eq:ConserveOracularInfo}
\end{align}
\item from conservation of cryptic and predictive information:
\begin{align}
\cInfo{\past{Z}}{R_0}{R_{k} \future{Z}} + \cInfo{\past{Z};R_0}{Z_{1:k}}{R_k} \hspace{-13em} & \nonumber \\
& = \cInfo{R_0; R_{k}}{Z_{1:k}}{\past{Z}} + \cInfo{R_{k}}{Z_{1:k}}{R_0 \past{Z}}.
\label{eq:ConserveCryptEInfo}
\end{align}
\end{enumerate}
\begin{proof}
To simplify the notation in the proof, we label information quantities by their associated label in the diagram \cref{fig:MemoryUpdate} (see also \cref{app:Anatomy}).

{\bf i.} 
By conservation of gauge information: 
\begin{equation}
\alpha + A = I + A.
\end{equation}
Eliminating the persistent gauge information $A$, and translating the diagram regions back into their informational quantities, we recover eq.~\eqref{eq:ConserveGaugeInfo}.

{\bf ii.}
By conservation of oracular information:
\begin{equation}
C+E+\delta+\zeta = C+II.
\end{equation}

Recall that region $E$ is no longer oracular once $Z_{1:k}$ has been produced -- thus, although information pertinent to this word is in the memory at both times, it is only oracular at time $0$.
We can eliminate the persistent oracular information $C$, and translate back into information quantities to recover eq.~\eqref{eq:ConserveOracularInfo}.

{\bf iii.}
By conservation jointly of cryptic information and excess entropy (i.e.\ all the information in the memory visible from past outputs at times $0$ and $k$ respectively):
\begin{equation}
\beta + \gamma + \epsilon + B + D + F = B + D + E + F + III + IV + V.
\end{equation}
(Recall, the information of region $E$, although oracular at time $0$, is visible in the output pattern by time $k$.)
Regions $IV$, $V$ and $\epsilon$ are empty,
 and regions $B$, $D$, and $F$ appear on both sides. Hence:
\begin{equation}
\beta + \gamma = E + III
\end{equation}
Translating this into information quantities recovers eq.~\eqref{eq:ConserveCryptEInfo}.
\end{proof}
\end{lemma}

{\bf Proof of Theorem 1.} {\em
For a model that generates $k$ steps of a pattern $\pastfuture{Y}$,
 the minimum dissipative cost of generation is bounded by the {\em discarded cryptic information} in the model's memory $R$:
\begin{align}
W_{\rm diss}^k = \kB T \; \cInfo{\past{Y}}{R_0}{\future{Y}R_{k}}.
\end{align}
}
\begin{proof}
Recall \cref{eq:GenDiss}:
\begin{align}
\frac{1}{\kB T} W & = \cEnt{R_0}{Y_{1:k} R_{k}} - \cEnt{R_{k}}{Y_{1:k} R_0} + \zeta_R\!\left(k\right).
\end{align}
This expression can be seen in \cref{fig:MemoryUpdate} as the difference between the blue ($\alpha + \beta + \zeta$) and yellow ($E+\delta$) regions and the red region ($I+II$).
That is,
\begin{equation}
\frac{1}{\kB T}  W_{\rm diss}^k = \alpha + \beta + \zeta + E + \delta - I - II.
\end{equation}
From lemma~\ref{lem:cons}i, we have $\alpha = I$, and from lemma~\ref{lem:cons}ii, $E+\delta+\zeta = II$, and hence the only remaining term is 
\begin{equation}
\frac{1}{\kB T} W_{\rm diss}^k = \beta.
\end{equation}
Translating ``$\beta$'' back into an information expression yields the claim.
\end{proof}

\section{The delay-buffer generator}
\label{app:DelayBufGen}

Before we begin our analysis of the delay--buffer generator,
 it is helpful to repeat one of the definitions of the cryptic order presented in \citet{MahoneyEJC11}:
\begin{definition}[Cryptic order]
\label{def:cryptic}
For a stationary pattern $\pastfuture{X}$ with causal states $S$, the {\bf cryptic order} is
\begin{align}
\label{eq:def:cryptic}
& k = \mathrm{min} \left\{ L \in \mathbb{Z}^+ : \cEnt{X_{L+1}}{S_0 X_{1:L}} \right.\nonumber \\
& \hspace{5em} \left. = \cEnt{X_0}{X_{1:L} S_{L}} \right\},
\end{align}
or is $\infty$ if no finite minimum can be found.
\end{definition}
Colloquially (at least to a computational mechanist!), since $\cEnt{X_{L+1}}{S_0 X_{1:L}} = \cEnt{X_{L+1}}{S_L} = \cEnt{X_1}{S_0}$,
 we can understand this quantity as the minimum size of the preceding word that a forward-predicting causal state must be augmented with to make a memory that is as effective at {\em retrodicting} its past as it is at predicting its future.
Equivalently, the cryptic order is the lowest $k \in \mathcal{Z}^+$ that satisfies $\cEnt{S_{k}}{X_{1:k} \future{X}_k} = 0$.
Since the Markov order is the lowest $m \in \mathcal{Z}^+$ such that $\cEnt{S_{m}}{X_{1:m}} = 0$, it is clear that a pattern's cryptic order will never be greater than its Markov order.

\subsection{Example mechanism}
\label{app:DBM}

\begin{figure}[hbt]
\includegraphics[width=0.45\textwidth]{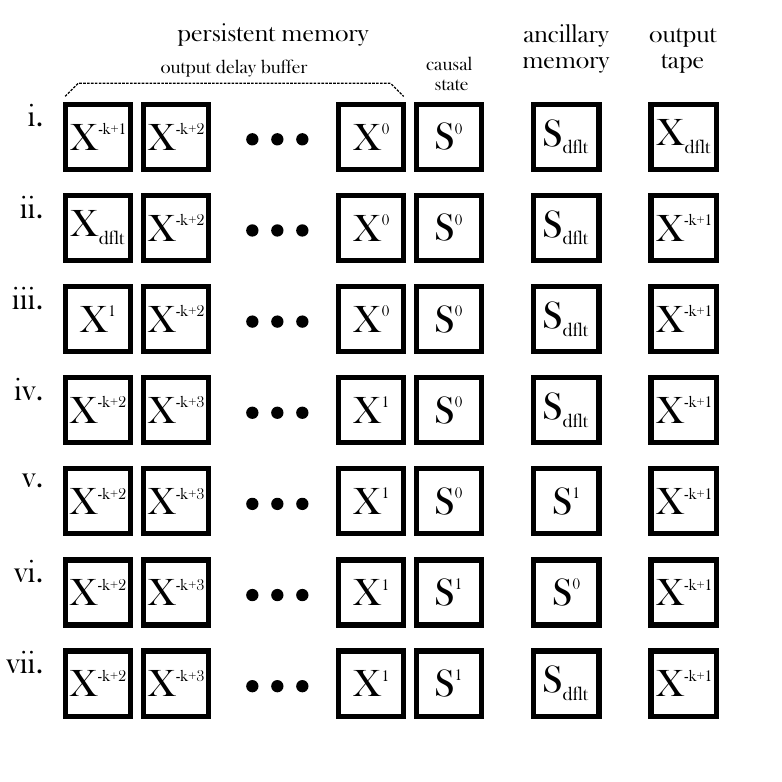}
\caption{
\label{fig:DelayBuffer}
\caphead{The delay-buffer generator.}
An $\epsilon$-machine is augmented with a delay buffer defers its by $k$ steps.
When $k$ matchs or exceeds the cryptic order, the Landauer minimum bound on generation cost matches the change in entropy rate of the output tape $\cEnt{X_1}{S_0} - \Ent{X_{\rm dflt}}$.
}
\end{figure}

A possible implementation of the delay buffer (operating with delay at or longer than cryptic order) is as follows (see sketch in \cref{fig:DelayBuffer}):-
\begin{enumerate}[label=\roman*.]
\item The machine begins in a memory state $X_{-k+1}\ldots X_0 S_{0}$, and has available to it a (pure) ancillary state $S_{\rm dflt}$ of the same dimensionality of causal state. 
A system on the tape (which will ultimately store the output) is inserted, initially in state $X_{\rm dflt}$.
\item The part of the memory containing $X_{-k-1}$ is reversibly swapped with the tape system.
{\em The output tape now has its correct final statistics.}
\item At work cost proportional to the difference between the entropy rate of the default state and the pattern (proven in Lemma~\ref{thm:DelayBuffer}), $\Ent{X_{\rm dflt}} - \cEnt{X_{1}}{S_0}$, the just-swapped portion of the memory is adjusted from $X_{\rm dflt}$ to $X_{1}$.
{\em This is the only heat-producing step.}
\item The buffer in the memory is (reversibly) cyclically shifted such that it now ranges from $X_{-k+2}$ to $X_{1}$.
\item Using $X_{1}$ and $S_0$ from within the memory, the ancillary system is reversibly changed from $S_{\rm dflt}$ to $S_{1}$ (causal states are unifilar\footnote{Unifilarity is the condition $\cEnt{R_1}{R_0 X_1} = 0$, i.e.\ if the previous internal state is known, then every output completely identifies the next internal state. In terms of state machine--diagrams: for each state, every arrow out of that particular state labelled by the same symbol will point to the same target state. $\varepsilon$--machines always have this property~\cite{ShaliziC01}.}; even if the memory as a whole is not). 
\item The ancillary system is reversibly swapped with the causal state part of the memory. {\em Every index in the main memory has now advanced by $1$, and the memory has updated from $R_{-k}$ to $R_{-k+1}$.}
\item To complete the generation, the ancillary system must be reset from $S_0$ back to its default state $S_{\rm dflt}$. 
However, with the available information in the generator, this can be done reversibly, since $\cEnt{S_{0}}{S_{1} X_{-k+1:0} X_1} = 0$ (from Lemma~\cref{lem:CrypticNoDiss}, below).
\end{enumerate}
Thus, a step of the pattern has been emitted and the memory has been updated, at total work cost $\Ent{X_{\rm dflt}} - \cEnt{X_{1}}{S_0}$.

Since this machine is already without dissipation, a generator with word length $m$ can be trivially realized by repeating the above process $m$ times, incurring a work cost proportional to the total change in entropy rate.

\subsection{Algorithm for finding the dynamics of the delay--buffer generator}
\label{sec:DBalgo}
The following algorithm can be used to list the dynamics (e.g.\ as in \cref{fig:DB_PC}) of a $k$--step delay buffer,
 representing it as an edge-emitting Hidden Markov Model (eeHMM).
 
\noindent {\bf Inputs: }
\begin{enumerate}
\item $\pastfuture{X}$ -- a stationary stochastic process over alphabet $\mathcal{X} = \{x_1, \ldots x_N\}$.
\item $k$ -- a non-negative integer denoting the desired delay length.
\end{enumerate}
{\bf Outputs:} 
\begin{enumerate}
\item $\mathcal{N}_{\rm out}$ -- a list of nodes of the eeHMM.
\item $\mathcal{E}_{\rm out}$ -- a list of weighted, labelled, edges, describing the dynamics of the eeHMM.
\end{enumerate}
{\bf Algorithm:}
\begin{enumerate}
\item Determine the $\varepsilon$--machine of the process $\pastfuture{X}$ (e.g.\ via \cite{CrutchfieldY89,ShaliziC01}). 
This provides a list {\em causal states} $\mathcal{S} := \{s_1, \ldots s_d\}$, 
 an encoding map $\varepsilon: \past{X} \to \mathcal{S}$,
 and a list $\mathcal{E}_{\varepsilon}$ of weighted, labelled transitions between causal states,
 where transition from $s_i\to s_j$ emitting symbol $x\in \mathcal{X}$ occurs with probability $p^{x}_{j|i}$,
\item If $k=0$ set $\mathcal{N}_{\rm out} = \mathcal{S}$ and $\mathcal{E}_{\rm out}=\mathcal{E}_{\varepsilon}$, then skip to step \ref{algo:end}.
\item Otherwise, for each causal state $s_i$ in $\mathcal{S}$:
\begin{enumerate}
\item For each $m$ denoting a choice of length $k$ sequence $x_{m_1} \ldots x_{m_k}$ in $\mathcal{X}^{\otimes k}$:
\begin{enumerate}
\item \label{algo:skipempty} If there are no semi-infinite sequences terminating with $x_{m_1} \ldots x_{m_k}$ such that $\varepsilon(\ldots x_{m_1} \ldots x_{m_k}) = s_i$, skip to the $m$ in the loop.
\item Otherwise, add state $s_i \otimes x_{m_1} \otimes x_{m_2} \otimes X_{m_3}$ to the list of output nodes $\mathcal{N}_{\rm out}$.
\item From the list $\mathcal{E}_{\varepsilon}$, for each edge beginning in $s_i$, and each pair $x,  j$ where $p^{x}_{j|i}>0$:
\begin{enumerate}
\item To the list of output edges, $\mathcal{E}_{\rm out}$, 
 add the edge
 from $s_i \otimes x_{m_1} \otimes  \ldots  \otimes  x_{m_k}$
 to $s_j \otimes x_{m_2} \otimes  \ldots \otimes  x_{m_k} \otimes x$,
 with label $x_{m_1}$,
 and weight $p^{x}_{j|i}$.
\end{enumerate}
\end{enumerate}
 \end{enumerate}
\item \label{algo:end}
$\mathcal{N}_{\rm out}$ and $\mathcal{E}_{\rm out}$ now describe the dynamics of the $k$-step delay buffer.
\end{enumerate}

By construction, all elements in $\mathcal{S}\otimes \mathcal{X}^{\otimes k}$ that are not in $\mathcal{N}_{\rm out}$ denote memory states that are unoccupied  in the stationary operation of the generator (corresponding to sequences in the buffer that would be incompatible with the memory's causal state).
As such, their exclusion has no impact on the analysis of the generator's entropic behaviour -- but allows for a minor amount of dimensional reduction in drawing the diagrams.

\noindent{\bf Example: The perturbed coin.}
The perturbed coin process models a binary system, whose state transitions at each time-step with probability $p$ (otherwise remaining the same). 
For instance, one could imagine a coin on a plate that is gently shaken at each time-step, and then recorded whether it is heads-up ($0$) or tails-up ($1$) following each shake.
The $\varepsilon$--machine for the perturbed coin (\cref{fig:DB_PC}a) can be described by causal states $s_0$, $s_1$ and the following directed edges:

\begin{center}
\begin{tabular}{ |c|c c| } 
\hline
Edge & Weight & Label \\
\hline
$s_0 \to s_0$ & 1-p & 0\\ 
$s_0 \to s_1$ & p & 1 \\  
$s_1 \to s_0$ & p & 0 \\
$s_1 \to s_1$ & 1-p & 1 \\
\hline
\end{tabular}
\end{center}

To produce the $k=1$ buffer (\cref{fig:DB_PC}b), going through the above algorithm generates the following edges:

\begin{center}
\begin{tabular}{ |c|c c| } 
\hline
Edge & Weight & Label \\
\hline
$s_0 \otimes 0 \to s_0 \otimes 0$  & $1-p$ & $0$ \\
$s_0 \otimes 0 \to s_1 \otimes 1$ & $p$ & $0$ \\
($s_0 \otimes 1$) & \multicolumn{2}{|l|}{Skipped -- no support.} \\
($s_1 \otimes 0$) & \multicolumn{2}{|l|}{Skipped -- no support.} \\
$s_1 \otimes 1 \to s_0 \otimes 0$ & $p$ & $1$ \\
$s_1 \otimes 1 \to s_1 \otimes 1$ & $1-p$ & $1$ \\
\hline
\end{tabular}
\end{center}
The nodes in brackets are skipped by step 3.a.i.\ of the algorithm, and so are omitted from $\mathcal{N}_{\rm out}$.
This is because the perturbed coin has no sequences ending in $\ldots 1$ that are mapped to causal state $s_0$, and likewise, no sequences ending in $\ldots 0$ that are mapped to causal state $s_1$.

Similarly, the $k=2$ buffer (\cref{fig:DB_PC}c), going through the above algorithm generates:

\begin{center}
\begin{tabular}{ |c|c c| } 
\hline
Edge & Weight & Label \\
\hline
$s_0 \otimes 0 \otimes 0 \to s_0 \otimes 0 \otimes 0$ &$1-p$ & $0$ \\
 $s_0 \otimes 0 \otimes 0 \to s_1 \otimes 0 \otimes 1$ &$p$ & $0$ \\
($s_0 \otimes 0 \otimes 1$) & \multicolumn{2}{|l|}{Skipped -- no support.} \\
$s_0 \otimes 1 \otimes 0 \to s_0 \otimes 0\otimes 0$ &$1-p$ & $1$ \\
 $s_0 \otimes 1 \otimes 0 \to s_1 \otimes 0\otimes 1$ &$p$ & $1$ \\
($s_0 \otimes 1 \otimes 1$) & \multicolumn{2}{|l|}{Skipped -- no support.} \\
($s_1 \otimes 0 \otimes 0$) & \multicolumn{2}{|l|}{Skipped -- no support.} \\
 $s_1 \otimes 0 \otimes 1 \to s_0 \otimes 1 \otimes 0$ &$p$ & $0$ \\
 $s_1 \otimes 0 \otimes 1 \to s_1 \otimes 1 \otimes 1$ &$1-p$ & $0$ \\
($s_1 \otimes 1 \otimes 0$) & \multicolumn{2}{|l|}{Skipped -- no support.} \\
 $s_1 \otimes 1 \otimes 1 \to s_0 \otimes 1\otimes 0$ &$p$ & $1$ \\
 $s_1 \otimes 1 \otimes 1 \to s_1 \otimes 1\otimes 1$ &$1-p$ & $1$ \\
 \hline
\end{tabular}
\end{center}

It can be seen that (for any delay $k\geq1$), each node only outputs one distinct symbol when it transitions.
Indeed, knowing the memory state gives perfect knowledge about the next $k$ steps of the pattern, reflecting the delay buffer's {\em oracular information}.
However, each node also transitions to multiple distinct target nodes that cannot be differentiated by this label alone,
 demonstrating that this machine is intrinsically non-unifilar -- a property that holds for any oracular machine.

\subsection{DBGs of finite cryptic order patterns}
\label{app:DelayBufGenDiss}

We prove the following entropic statement:
\begin{lemma}
\label{lem:CrypticNoDiss}
For a stationary pattern $\pastfuture{X}$ with causal states $S$,
\begin{align}
\cEnt{S_{0}}{X_{-k+1:0} X_{1}S_{1}} = 0,
\end{align}
when $k$ is greater than or equal to the cryptic order of $\pastfuture{X}$.
\begin{proof}
Consider the joint entropy of $X_{-k+1:0} S_0 X_{1} S_{1}$, expanded in two ways:
\begin{align}
\Ent{X_{-k+1:0} S_0 X_{1} S_{1}} 
& = \Ent{X_{-k+1:0}} + \cEnt{S_0}{X_{-k+1:0}} \nonumber \\
& \qquad + \cEnt{S_{1}X_{1}}{S_0 X_{-k+1:0}} \nonumber \\
& = \Ent{X_{-k+2:0}X_1} + \cEnt{S_{1}}{X_{-k+2:1}} \nonumber \\
& \qquad + \cEnt{S_{0}X_{-k+1}}{S_{1} X_{-k+2:1}}.
\end{align}
From stationarity, the first two terms of each expansion are equal, and hence:
\begin{align}
\label{eq:WeirdOffset}
\cEnt{S_{1}X_{1}}{S_0 X_{-k+1:0}} & = \cEnt{S_{0}X_{-k+1}}{S_{1} X_{-k+2:0} X_1}.
\end{align}
We can then expand the left-hand-side:
\begin{align}
\cEnt{S_{1}X_{1}}{S_0 X_{-k+1:0}} \hspace{-6.5em}& \nonumber \\
&  = \cEnt{X_{1}}{S_0 X_{-k+1:0}} + \cEnt{S_{1}}{S_0 X_{-k+1:0}  X_1 } \nonumber \\
& = \cEnt{X_{1}}{S_0 X_{-k+1:0} } = \cEnt{X_{1}}{S_0} \nonumber \\
& = \cEnt{X_{k+1}}{S_0 X_{1:k}},
\end{align}
where we have used the unifilarity of causal states to set $0 \leq \cEnt{S_{1}}{S_0 X_{-k+1:0} X_1} \leq \cEnt{S_{1}}{S_0 X_1} =0$ eliminating the second term,
 and the property of {\em causal shielding} to simplify the remaining expression (conditioning on additional $X_{t \leq 0}$ in the past of $S_0$ cannot improve any predictions about future $X_{t>0}$), and then unifilarity and stationarity in the final equality.
We also expand the right hand side of \cref{eq:WeirdOffset}
\begin{align}
\cEnt{S_{0}X_{-k+1}}{ X_{-k+2:1} S_1} \hspace{-10em} &  \nonumber \\
& = \cEnt{X_{-k+1}}{X_{-k+2:1}S_1} + \cEnt{S_{0}}{X_{-k+1:0} X_1 S_1} \nonumber \\
& = \cEnt{X_{0}}{X_{1:k}S_k} + \cEnt{S_{0}}{X_{-k+1:0} X_1 S_1}.
\end{align}
Substituting these expressions back into \cref{eq:WeirdOffset} yields
\begin{align}
\cEnt{S_{0}}{S_{1} X_{-k+1:0} X_1} \hspace{-10em} & \nonumber \\
& = \cEnt{X_{k+1}}{S_0 X_{1:k}} - \cEnt{X_{0}}{X_{1:k} S_{k}}.
\label{eq:EqualCrypt}
\end{align}
This difference is exactly the two terms that must be equated in the definition of the cryptic order (\cref{def:cryptic}).
Hence, if $k\geq L$, where $L$ is the cryptic order, these two terms are equal and thus
\begin{align}
\cEnt{S_{0}}{X_{-k+1:0} X_1 S_{1}} = 0 \qquad k\geq L.
\end{align}
\end{proof}
\end{lemma}

\begin{lemma}
\label{thm:DelayBuffer}
For any pattern $\pastfuture{X}$ with finite cryptic order,
 there is a finite-memory generator for every word length $L$ with $W^L_{\rm diss} = 0$.
\begin{proof}
Proof is by construction of the cryptic order DBG.
Let the alphabet of the pattern be $\mathcal{X}$, and of the causal states be $\mathcal{S}$,
 and write the DBG memory as $\mathcal{R} = \mathcal{X}^{\otimes k} \otimes \mathcal{S}$,
 where $k$ is the cryptic order of $\pastfuture{X}$.
In particular, the state of the memory $R_0$ at time $-k$ is explicitly:
\begin{equation}
R_{-k} = X_{-k+1:0} S_0.
\end{equation}
That is, the memory is composed of a causal state $S_0$ augmented by a sequence of $k$ steps of the pattern $X_{-k+1}\ldots X_0$ that immediately precede $S_0$.

Let us consider the entropic changes manifest by running this generator.
In particular, we start from a state $R_{-k}$ and the output tape in state $X_{\rm dflt}$, and finish with the memory in state $R_{-k+1}$ and the output tape in state $X_{-k+1}$.
From Landauer's principle, the minimum work cost is proportional to the difference in entropy:
\begin{equation}
\beta W = \left[ \Ent{R_{-k} X_{\rm dflt}} - \Ent{R_{-k+1} X_{-k+1}} \right].
\end{equation}
Noting that $X_{\rm dflt}$ and $R_{-k}$ are totally uncorrelated, we expand the above substituting in the explicit form of the memory $R$:
\begin{align}
\beta W & = \left[\Ent{X_{\rm dflt}} +  \Ent{X_{-k+1:0} S_0} \right.  \nonumber\\
& \quad - \left. \Ent{X_{-k+1} X_{-k+2:1} S_{1}} \right].
\end{align}
Now consider expanding in two ways:
\begin{align}
\Ent{X_{-k+1:0} S_0 X_{1} S_{1} } \hspace{-10em} & \nonumber \\
& = \Ent{X_{-k+1:0} S_0} + \cEnt{ X_{1} S_{1} }{X_{-k+1:0} S_0} \nonumber \\
& = \Ent{X_{-k+1:0} X_{1} S_{1}} + \cEnt{ S_{0} }{X_{-k+1:0} X_{1} S_{1}},
\end{align}
such that
\begin{align}
\Ent{X_{-k+1:0} S_0}  - \Ent{X_{-k+1} X_{-k+2:1} S_{1}} \hspace{-15em} \nonumber \\
 & = \cEnt{ S_0 }{X_{-k+1:0} X_{1} S_{1}}  -  \cEnt{ X_{1} S_{1} }{X_{-k+1:0} S_0} \nonumber \\
& = -\cEnt{X_1}{S_0},
\end{align}
where we have used Lemma~\cref{lem:CrypticNoDiss} to set the first term to $0$, 
 and the causal shielding and unifilar properties of causal states to simplify the second term.

It then follows
\begin{equation}
\beta W = \Ent{X_{\rm dflt}} - \cEnt{X_{1}}{S_{0}} = W_{\rm tape}
\end{equation}
and $W^1_{\rm mem} = 0$.
Since this cost is zero, the update can be repeated $L$ times to produce a machine with $W^L_{\rm diss} = 0$ for all $L\geq 1$.
\end{proof}
\end{lemma}

\subsection{DBGs of infinite cryptic order patterns}
\label{app:InfiniteCryptic}
By imposing a long enough delay the dissipation associated with generating any pattern with a finite number of causal states goes to zero -- even if that pattern has infinite cryptic order.

\begin{lemma}
\label{lem:ArbLow}
Let $\pastfuture{X}$ be some stationary pattern with a finite number of causal states.
There for any $\delta>0$, there exists a finite $L$ such that $\cEnt{S_L}{X_{0:L}} < \delta$.
\begin{proof}
Travers and Crutchfield~\cite{TraversC10,TraversC11} show that for {\em any} $\epsilon$-machine with a finite number of causal states,
 not only does $\lim_{L\to\infty} \cEnt{S_L}{X_{0:L}} \to 0$, but this is a pointwise exponential convergence.
It immediately follows that for any $\delta>0$, a sufficiently long $L$ can be found such that  $\cEnt{S_L}{X_{0:L}}$ is strictly less than $\delta$.
\end{proof}
\end{lemma}
I will outline a few points for the reader's intuition, but strongly suggest they refer to the citations~\cite{TraversC10,TraversC11} for mathematical detail.
First, if the machine has a finite Markov order, $K$, one can simply choose $L\geq K$ and then $\cEnt{S_L}{X_{0:L}}=0 < \delta$.
Second, if the machine has a finite length synchronizing word of length $L'$ (such that after observing this word, the causal state then known with certainty),
 then for $L>L'$, as $L$ increases, the probability of observing this synchronizing word tends to unity,
  and the entropy accordingly decreases to $0$.
These two cases are known as {\em exactly synchronizing} machines~\cite{TraversC10}.

\begin{figure}[hbt]
\includegraphics[width=0.3\textwidth]{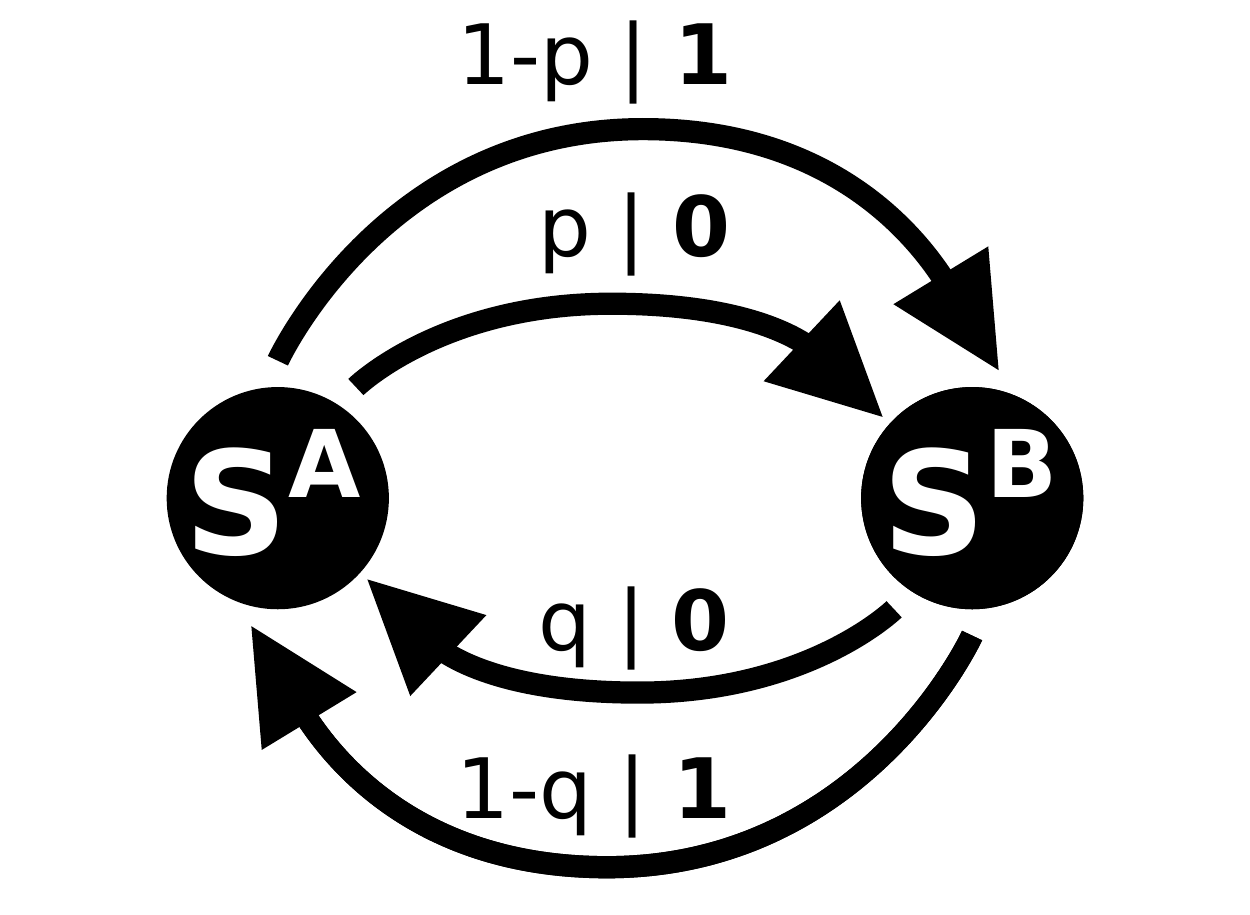}
\caption{
\label{fig:ABC}
\caphead{Example: Alternating biased coin.}
No finite length sequence of $0$s and $1$s will identify the causal state of this process with certainty.
Nonetheless, when $p\neq q$, the longer the observed sequence, the more certainty we have about the state of the machine: a property known as {\em asymptotic synchronization}.
}
\end{figure}

The remaining case -- {\em strictly asymptotic synchronization}~\cite{TraversC11} -- admit no such finite synchronizing words. 
For example: consider the so-called ``alternating biased coin'' process, with two causal states $S^A$ and $S^B$ (\cref{fig:ABC}). 
In $S^A$, there is probability $p$ of emitting 0 and $1-p$ of emitting 1, transferring in both cases to the other state $S^B$, which has probability $q\neq p$  (resp.\ $1-q$) of emitting 0 (resp.\ 1) before transitioning back.
Since {\em all} binary sequences are permissible, whether the machine started in $S^A$ or $S^B$, no finite-length sequence can identify the causal state with perfect certainty.

However, crucially, all patterns with a finite number of causal states are (at least) asymptotically synchronizing: 
 the definition of causal states requires different states to have divergent statistics (observable after a finite sequence for finite-state machines).
Also, due to the unifilarity of $\epsilon$-machines, on average one never becomes {\em less} certain about the causal state through the observation of longer sequences.
Then, the observation of ever-longer strings effectively amounts to hypothesis--testing over ever--larger samples whether the sequence began in a particular causal state.

\begin{lemma}
\label{lem:ArbitraryLowDissipation}
For any pattern $\pastfuture{X}$ with a finite number of causal states,
 and positive work value $\epsilon > 0$,
 there is a finite-memory generator for every word length $k$ with $W^k_{\rm diss} < \epsilon$.
\begin{proof}
Consider a $l$-step causal-state delay-buffer machine (as above) with memory $R_0 = X_{1:l} S_{l}$.
Recall from Lemma~\ref{thm:CrypticDissipation} that the minimum dissipation $W^k_{\rm diss}$ is proportional to 
\begin{align}
\cInfo{\past{X}}{{R_0}}{\future{X}R_{k}} & = \cInfo{\past{X}}{{R_0}}{\future{X}R_{k}} \nonumber \\
& = \cInfo{\past{X}}{{X_{1:l} S_{l}}}{\future{X}X_{k+1:k+l} S_{l+k}} \nonumber \\
& = \cInfo{\past{X}}{{S_{l}}}{\future{X}S_{l+k}}.
\end{align}
In the second line, we have eliminated repeated variables since $X_{k+1:k+l} \subset \future{X}$, and used $\cInfo{A}{BC}{CD} = \cEnt{BC}{CD}-\cEnt{BC}{ACD} = \cEnt{B}{D}-\cEnt{B}{AD} = \cInfo{A}{B}{D}$ to eliminate $X_{1:l}$.

Consider then:
\begin{align}
\label{eq:II1}
\iInfo{\past{X}}{S_{l}}{\future{X}S_{l+k}} & = \Info{\past{X}}{S_{l}} - \cInfo{\past{X}}{S_{l}}{\future{X}{S_{l+k}}} 
\end{align}
and
\begin{align}
\label{eq:II2}
\iInfo{\past{X}}{S_{l}}{\future{X}S_{l+k}}
& = \Info{S_{l}}{\future{X}S_{l+k}} - \cInfo{S_{l}}{\future{X}S_{l+k}}{\past{X}}  \nonumber \\
& = \Info{S_{l}}{\future{X}S_{l+k}} - \cEnt{S_{l}}{\past{X}},
\end{align}
where we have used
\begin{align}
\cInfo{S_{l}}{\future{X}S_{l+k}}{\past{X}} & = \cEnt{S_{l}}{\past{X}} - \cEnt{S_{l}}{\future{X}S_{l+k}\past{X}} \nonumber\\
& = \cEnt{S_{l}}{\past{X}},
\end{align}
noting that the second term in the top line is zero, as it conditions a causal state on the entire pattern and hence can be perfectly determined (by virtue of every pattern being asymptotically synchronizable).

Equating \cref{eq:II1,eq:II2} gives:
\begin{align}
\cInfo{\past{X}}{S_{l}}{\future{X}{S_{l+k}}} \hspace{-4em} & \nonumber \\
& = \cEnt{S_{l}}{\past{X}} + \Info{\past{X}}{S_{l}} - \Info{S_{l}}{\future{X}S_{l+k}} \nonumber \\
& = \cEnt{S_{l}}{\past{X}} + \Ent{S_{l}} - \cEnt{S_{l}}{\past{X}} \nonumber \\
& \qquad - \Ent{S_{l}} + \cEnt{S_{l}}{\future{X} S_{l+k}} \nonumber \\
& = \cEnt{S_{l}}{\future{X} S_{l+k}}.
\end{align}

However $\cEnt{S_{l}}{\future{X} S_{l+k}} \leq \cEnt{S_{l}}{X_{1:l+k}}$ since $X_{1:l+k} \subset \future{X}$,
 and by Lemma~\ref{lem:ArbLow} for arbitrary $\epsilon > 0$, $\cEnt{S_{l}}{X_{0:l+k}} < \epsilon$ for some large enough $l+k$.
Hence, the dissipation can be made arbitrarily small by choosing a sufficiently long, but finite, delay.
\end{proof}
\end{lemma}

\vspace*{1em}
\section{Thermodynamics of forecasting}
\label{app:Forecast}
\begin{lemma}
\label{eq:ThermoForecast}
For a forecaster with generic memory $R$ that follows $k$ steps of a pattern $\pastfuture{X}$,
 the minimum work cost $W$ is bounded by:
\begin{align}
\frac{1}{\kB T} W = \cInfo{\past{X}}{R_0}{\future{X}R_{k}} - \zeta_R(k).
\end{align}
\begin{proof}
Recall \cref{eq:ForecastEntropy}:
\begin{align}
\Delta H & = \Info{R_0}{X_{1:k}} - \Info{R_k}{X_{1:k}}.
\end{align}
Using the information diagram (\cref{app:Anatomy}, \cref{fig:MemoryUpdate}) we express this as
\begin{align}
\Delta H & = \left(E + F + \gamma + \delta\right) - \left(E + F + III \right) \nonumber  \\
& = \gamma + \delta - III.
\end{align}
Lemma \ref{lem:cons}(iii) states $\beta + \gamma = E - III$ and hence
\begin{align}
\label{eq:ForecastCost}
\Delta H & = E + \delta - \beta.
\end{align}
``$\beta$'' corresponds to the discarded cryptic information $\cInfo{\past{X}}{R_0}{\future{X}R_{k}}$.
Meanwhile, ``$E + \delta$'' corresponds to $\cInfo{R_0}{X_{1:k}}{\past{X}} =: \zeta_R(k)$, the oracular information about the word $X_{1:k}$.
Inserting these terms into \cref{eq:ForecastCost} and applying Landauer's principle proves the claim.
\end{proof}
\end{lemma}

\end{document}